\documentclass[11pt]{article}
\usepackage{amsmath,amssymb,bbm,amsthm}
\usepackage{fullpage,hyperref,xspace,thm-restate,thmtools}
\usepackage{cleveref,color,nicefrac}
\usepackage{graphicx,tikz,pgfplots}
\usepackage{wrapfig}
\usetikzlibrary{patterns}

\usepackage[bottom=1in,top=1in,left=1in,right=1in]{geometry}
\usepackage[compact]{titlesec}

\usepackage{setspace}

\declaretheorem[numberwithin=section]{theorem}
\declaretheorem[sibling=theorem]{lemma}
\declaretheorem[sibling=theorem]{claim}

\declaretheorem[sibling=theorem]{corollary}
\declaretheorem[sibling=theorem]{definition}
\declaretheorem[sibling=theorem]{assumption}

\makeatletter
\newcounter{algocounter}
\@ifpackageloaded{hyperref}%
  {\newcommand{\mylabel}[2]
    {\refstepcounter{algocounter}\protected@write\@auxout{}{\string\newlabel{#1}{{\textcolor{blue}{\textup{#2}}}{\thepage}%
      {\@currentlabelname}{\@currentHref}{}}}}}%
\makeatother

\usepackage{framed,xcolor}
\renewenvironment{leftbar}[1][\hsize]
{%
    \MakeFramed{\hsize#1\advance\hsize-\width\FrameRestore}%
}
{\endMakeFramed}
\newcommand{\BLEFT}{\begin{leftbar}}
\newcommand{\ELEFT}{\end{leftbar}}

\newenvironment{subproof}[1][\proofname]{%
  \begin{proof}[#1]%
}{%
  \end{proof}%
}

\usepackage{algorithm}
\usepackage[noend]{algpseudocode}
\algnewcommand{\IIf}[1]{\State\algorithmicif\ #1\ \algorithmicthen}
\algnewcommand{\EndIIf}{\unskip\ \algorithmicend\ \algorithmicif}
\algrenewcommand\algorithmiccomment[2][\normalsize]{{#1\hfill\(\triangleright\) \emph{#2}}}

\usepackage[colorinlistoftodos,textsize=tiny,textwidth=2cm,color=red!25!white,obeyFinal]{todonotes}
\newcommand{\agnote}[1]{\todo[color=blue!25!white]{AG: #1}\xspace}
\newcommand{\elnote}[1]{\todo[color=green!25!white]{EL: #1}\xspace}
\newcommand{\jlnote}[1]{\todo[color=red!25!white]{JL: #1}\xspace}

\begin{document}

\newcommand{\f}{\frac}
\newcommand{\cd}{\cdot}
\newcommand{\bn}{\binom}
\newcommand{\sr}{\sqrt}
\newcommand{\cds}{\cdots}
\newcommand{\lds}{\ldots}
\newcommand{\vds}{\vdots}
\newcommand{\dds}{\ddots}
\newcommand{\pge}{\succeq}
\newcommand{\ple}{\preceq}
\newcommand{\bs}{\backslash}
\newcommand{\s}{\subseteq}
\newcommand{\ol}{\overline}

\newcommand{\BE}{\begin{enumerate}}
\newcommand{\EE}{\end{enumerate}}
\newcommand{\BOn}{\begin{OneLiners}}
\newcommand{\EOn}{\end{OneLiners}}
\newcommand{\im}{\item}
\newcommand{\BI}{\begin{itemize}}
\newcommand{\EI}{\end{itemize}}

\newcommand{\Sum}{\displaystyle\sum\limits}
\newcommand{\Prod}{\displaystyle\prod\limits}
\newcommand{\Int}{\displaystyle\int\limits}
\newcommand{\Lim}{\displaystyle\lim\limits}
\newcommand{\Max}{\displaystyle\max\limits}
\newcommand{\Min}{\displaystyle\min\limits}
\renewcommand{\Cup}{\displaystyle\bigcup\limits}

\newcommand{\logn}{\log n}

\newcommand{\dx}{\frac d{dx}}
\newcommand{\dy}{\frac d{dy}}
\newcommand{\dz}{\frac d{dz}}
\newcommand{\dt}{\frac d{dt}}

\newcommand{\inv}{^{-1}}

\newcommand{\R}{\mathbb R}
\newcommand{\Z}{\mathbb Z}
\newcommand{\F}{\mathbb F}
\newcommand{\C}{\mathbb C}
\newcommand{\N}{\mathbb N}
\newcommand{\Q}{\mathbb Q}

\newcommand{\eps}{\varepsilon}
\newcommand{\e}{\varepsilon}
\newcommand{\de}{\delta}
\newcommand{\De}{\Delta}
\newcommand{\la}{\lambda}
\newcommand{\g}{\gamma}
\newcommand{\G}{\Gamma}
\newcommand{\pt}{\partial}
\newcommand{\al}{\alpha}
\newcommand{\be}{\beta}
\newcommand{\om}{\omega}
\newcommand{\Om}{\Omega}
\newcommand{\el}{\ell}
\renewcommand{\th}{\theta}
\newcommand{\Th}{\Theta}
\newcommand{\m}{\mathcal}

\newcommand{\ba}{\mathbf a}
\newcommand{\bb}{\mathbf b}
\newcommand{\bc}{\mathbf c}
\renewcommand{\bf}{\mathbf f}
\newcommand{\bu}{\mathbf u}
\newcommand{\bv}{\mathbf v}
\newcommand{\bx}{\mathbf x}
\newcommand{\by}{\mathbf y}
\newcommand{\bz}{\mathbf z}

\newcommand{\Ra}{\Rightarrow}

\newcommand{\lf}{\lfloor}
\newcommand{\rf}{\rfloor}
\newcommand{\lc}{\lceil}
\newcommand{\rc}{\rceil}

\newcommand{\E}{\mathbb E}
\newcommand{\Var}{\text{Var}}
\newcommand{\Cov}{\text{Cov}}
\newcommand{\1}{\mathbbm 1}
\newcommand{\poly}{\text{poly}}
\newcommand{\polylog}{\text{polylog}}
\newcommand{\norm}[1]{\left\lVert#1\right\rVert}

\newcommand{\rank}{\text{rank}}
\newcommand{\spn}{\text{span}}
\newcommand{\Tr}{\text{Tr}}

\newcommand{\lp}{\left(}
\newcommand{\rp}{\right)}
\newcommand{\lb}{\left[}
\newcommand{\rb}{\right]}
\newcommand{\lmt}{\left[\begin{matrix}}
\newcommand{\rmt}{\end{matrix}\right]}

\newcommand{\BT}{\begin{theorem}}
\newcommand{\ET}{\end{theorem}}
\newcommand{\BL}{\begin{lemma}}
\newcommand{\EL}{\end{lemma}}
\newcommand{\BD}{\begin{definition}}
\newcommand{\ED}{\end{definition}}
\newcommand{\BC}{\begin{corollary}}
\newcommand{\EC}{\end{corollary}}
\newcommand{\BO}{\begin{observation}}
\newcommand{\EO}{\end{observation}}
\newcommand{\BCL}{\begin{claim}}
\newcommand{\ECL}{\end{claim}}
\newcommand{\BP}{\begin{proof}}
\newcommand{\EP}{\end{proof}}
\newcommand{\BPS}{\begin{proof}[Proof (Sketch)]}
\newcommand{\EPS}{\end{proof}}

\newcommand{\para}{\paragraph}
\newcommand{\defn}{\textbf}
\newcommand{\alert}{\textcolor{red}}
\newcommand{\blue}{\textcolor{blue}}

\newcommand{\Ttree}{\text{T-tree}\xspace}
\newcommand{\wbar}{\bar{w}}
\newcommand{\OPTbar}{\overline{\small OPT}}

\newcommand{\SMinKCut}{\ref{SimpleMinKCut}\xspace}
\newcommand{\MinKCut}{\ref{MinKCut}\xspace}
\newcommand{\Enum}{\ref{EnumCuts}\xspace}

\newcommand{\Bad}{\textsf{Bad}}
\newcommand{\thr}{\ensuremath{2^{[3]}\bs\{\emptyset,[3]\}}\xspace}
\newcommand{\Thr}{\ensuremath{(3,2^{[3]}\bs\{\emptyset,[3]\})}\xspace}
\newcommand{\Out}{\textup{out}}
\newcommand{\In}{\textup{in}}
\newcommand{\Cont}{\textup{cont}}
\newcommand{\Small}{\textup{small}}
\newcommand{\Largee}{\textup{large}}
\newcommand{\calR}{\mathcal{R}}
\newcommand{\calA}{\mathcal{A}}

\newcommand{\tsty}{\textstyle}

\newcommand{\Rone}{\m R_{\textup{one}}}
\newcommand{\Rboth}{\m R_{\textup{both}}}
\newcommand{\three}{\ensuremath{2^{[3]}}\xspace}
\newcommand{\Three}{\ensuremath{(3,2^{[3]})}\xspace}
\newcommand{\thre}{\ensuremath{2^{[3]}\bs\{\emptyset\}}\xspace}
\newcommand{\Thre}{\ensuremath{(3,2^{[3]}\bs\{\emptyset\})}\xspace}

\newcounter{myLISTctr}
\newcommand{\initOneLiners}{%
    \setlength{\itemsep}{0pt}
    \setlength{\parsep }{0pt}
    \setlength{\topsep }{0pt}
}
\newenvironment{OneLiners}[1][\ensuremath{\bullet}]
    {\begin{list}
        {#1}
        {\initOneLiners}}
    {\end{list}}

\newcommand{\Conn}{\kappa}
\renewcommand{\emptyset}{\varnothing}
\newcommand{\fEN}{2^k}

\newcommand{\kcut}{\ensuremath{k\textsc{-Cut}}\xspace}
\newcommand{\kkcut}{\ensuremath{(k-1)\textsc{-Cut}}\xspace} 
\newcommand{\tO}{\widetilde{O}}
\newcommand{\kclique}{\textsc{Max-Weight~\ensuremath{k}\textsc{-Clique}}\xspace}
\newcommand{\kkclique}{\textsc{Max-Weight~\ensuremath{(k-1)}\textsc{-Clique}}\xspace}
\newcommand{\ukclique}{\textsc{Unweighted~\ensuremath{(k-1)}\textsc{-Clique}}\xspace}
\newcommand{\ukkclique}{\textsc{Unweighted~\ensuremath{k}\textsc{-Clique}}\xspace}
\newcommand{\Opt}{\ensuremath{\mathsf{Opt}}\xspace}

\newcommand{\calS}{\mathcal{S}}
\newcommand{\calP}{\mathcal{P}}

\newcommand{\TIME}{1.981}

\title{\textbf{The Number of Minimum $k$-Cuts: \\ Improving the Karger-Stein Bound}}
\author{ Anupam Gupta\thanks{{\tt anupamg@cs.cmu.edu}. Supported in part by NSF awards CCF-1536002, CCF-1540541,
and CCF-1617790, and the Indo-US Joint Center for Algorithms Under Uncertainty. } \\ CMU \and Euiwoong Lee\thanks{{\tt euiwoong@cims.nyu.edu}. Supported in part by the Simons Collaboration on Algorithms and Geometry. }\\ NYU
  \and Jason Li\thanks{{\tt jmli@cs.cmu.edu}. Supported in part by NSF awards
    CCF-1536002, CCF-1540541, and CCF-1617790. }\\ CMU}
\date{}

\maketitle
\thispagestyle{empty}

\begin{abstract}
  Given an edge-weighted graph, how many minimum $k$-cuts can it have?
  This is a fundamental question in the intersection of algorithms,
  extremal combinatorics, and graph theory. It is particularly
  interesting in that the best known bounds are \emph{algorithmic}: they
  stem from algorithms that compute the minimum $k$-cut.

  In 1994, Karger and Stein obtained a randomized contraction algorithm
  that finds a minimum $k$-cut in $O(n^{(2-o(1))k})$ time. It can also
  \emph{enumerate} all such $k$-cuts in the same running time,
  establishing a corresponding extremal bound of
  $O(n^{(2-o(1))k})$. Since then, the algorithmic side of the minimum
  $k$-cut problem has seen much progress, leading to a deterministic
  algorithm based on a tree packing result of Thorup, which enumerates
  all minimum $k$-cuts in the same asymptotic running time, and gives an
  alternate proof of the $O(n^{(2-o(1))k})$ bound. However, beating the
  Karger--Stein bound, even for computing a single minimum $k$-cut, has
  remained out of reach.

  In this paper, we give an algorithm to enumerate all minimum
  $k$-cuts in $O(n^{(1.981+o(1))k})$ time, breaking 
  the algorithmic and extremal barriers for enumerating minimum
  $k$-cuts. To obtain our result, we combine ideas from both the
  Karger--Stein and Thorup results, and draw a novel connection between
  minimum $k$-cut and \emph{extremal set theory}. In particular, we give
  and use tighter bounds on the size of set systems with bounded dual
  VC-dimension, which may be of independent interest.
\end{abstract}

\newpage

\setcounter{page}{1}

\section{Introduction}
\label{sec:introduction}

We consider the \kcut problem: given an edge-weighted
graph $G = (V,E,w)$ and an integer $k$, delete a minimum-weight set of
edges so that $G$ has at least $k$ connected components. This problem is
a natural generalization of the global min-cut problem, where the goal
is to break the graph into $k=2$ pieces.  This problem has been actively
studied in theory of both exact and approximation algorithms, where each
result brought new insights and tools on graph cuts.

Goldschmidt and Hochbaum gave the first polynomial-time algorithm for
fixed $k$, with $O(n^{(1/2 - o(1))k^2})$
\footnote{$o(1)$ in the exponent indicates a quantity that goes to $0$ as $k$ increases. }
 runtime~\cite{GH94}. Since
then, the exact exponent in terms of $k$ has been actively studied.  The
textbook minimum cut algorithm of Karger and Stein~\cite{KS96}, based on
random edge contractions, can be adapted to solve \kcut in
$\tO(n^{2(k-1)})$ (randomized) time. The deterministic algorithms side
has seen a series of improvements since
then~\cite{KYN06,Thorup08,chekuri2018lp}. The fastest algorithm for
general edge weights is due to Chekuri et al.~\cite{chekuri2018lp}. It
runs in $O(mn^{2k-3})$ time and is based on a deterministic tree packing
result of Thorup~\cite{Thorup08}. Hence, the leading algorithms on the
randomized and deterministic fronts utilize completely different
approaches, but neither is able to break the $O(n^{(2-o(1))k})$ bound
for the problem.

On the lower bounds side, a simple reduction from $\kkclique$ to $\kcut$
implies that the conjectured time lower bound
$\tilde{\Omega}(n^{(1-o(1))k})$ for $\kkclique$~\cite{KClique} also
holds for $\kcut$.\footnote{The conjectured lower bounds are 
$\tilde{\Omega}(n^{(1-o(1))k})$ for \kclique when weights are integers in the range $[1,\Om(n^k)]$, and
$\tilde{\Omega}(n^{(\omega/3)k})$ for \ukclique, where $\omega$ is the
matrix multiplication constant. 
}
Given the recent interest in fine-grained complexity,
\emph{for the \underline{algorithmic} problem of finding the minimum $k$-cut, should the
true exponent in $n$ for \kcut be $k$, $2k$, or somewhere in between?}


Another closely related, extremal question concerns the number of
minimum $k$-cuts that a graph can have. The algorithms of Karger-Stein,
Thorup, and Chekuri et al.\ can be adapted to enumerate all minimum
$k$-cuts in $O(n^{(2-o(1))k})$ time, implying the same bound for the
extremal number of minimum $k$-cuts in a graph. To this date, no proof
of a better bound---algorithmic or otherwise---is known. On the other
hand, there are some graphs (e.g., a cycle with $n$ vertices) where the
number of minimum $k$-cuts is $\Omega(n^{k})$.\footnote{Technically, it
  is $\Om(n^k/k!)$, but we assume that $k$ is a large but fixed constant
  throughout.}  Thus, the mathematical question remains: \emph{for the
\underline{extremal} number of minimum $k$-cuts of a graph, should the exponent of
$n$ be $k$, $2k$, or somewhere in between?}

In recent work~\cite{GLL18focs}, we improved on the algorithmic barrier
of $O(n^{(2-o(1))k})$ for the case when the input graph has edge weights
that are integers polynomially bounded in $n$. In particular, our deterministic
algorithm runs in time $O(k^{O(k)} n^{(2\om/3+o(1))k})$, where
$\om \leq 2.3738$ denotes the matrix multiplication
constant~\cite{le2014powers,williams2012multiplying}. Aside from the unfortunate restriction to
integer-weighted graphs, our algorithm suffers other disadvantages
compared to the previous \kcut algorithms.

\begin{enumerate}
\item When edge weights are integers bounded by $n^{O(1)}$ (in
  particular, exponent independent of $k$), the reduction from \kkclique
  no longer holds. Consequently, our algorithm solves an easier variant
  of \kcut whose lower bound is only $\Om(n^{(\om/3-o(1))k})$ 
  from \ukkclique, and not $\Om(n^{(1-o(1))k})$.


\item The previous algorithms are ``combinatorial'', where ours uses
  fast matrix-multiplication as a
  black-box. (See~\cite{williams2010subcubic, abboud2015if} for
  discussion about combinatorial and matrix multiplication-based
  algorithms for \textsc{$k$-Clique} and other problems.) One concrete
  difference is that our algorithm inherently requires $n^{\Omega(k)}$
  space, whereas all of the aforementioned \kcut algorithms require only
  $\poly(n)$ space.

\item Our algorithm only finds \emph{one} minimum $k$-cut and, unlike
  the previous algorithms, cannot be adapted to find all of them in the
  same asymptotic running time. (This is a common weakness with
  algorithms that use matrix multiplication.) Hence, it does not imply
  any improved bound on the extremal number of minimum $k$-cuts, even
  for integer-weights.
\end{enumerate} 

To summarize, beating either of the $O(n^{(2-o(1))k})$ algorithmic and
extremal bounds has remained open for the general case. In this paper,
we break these bounds for large enough $k$, and get the first true
improvement over the Karger--Stein result for \kcut, in both the
algorithmic and extremal settings.


\BT[Enumeration Algorithm] \label{thm:main} For any large enough constant $k$, there is a randomized algorithm that enumerates all
minimum $k$-cuts in time $O(n^{\TIME k})$ w.h.p.  \ET

\BC[Extremal Result]\label{cor:main} For any large enough constant $k$, there are at most $O(n^{\TIME k})$ many minimum
$k$-cuts.  \EC


\subsection{Our Techniques}
\label{sec:techniques}

We believe that an important component of the paper's contributions are
the techniques, which draw on different areas: some of them are perhaps
surprising and suggesting directions for further
investigation. 
As mentioned previously, the two leading approaches so far to
\kcut---random contractions and tree packing---utilize completely
different techniques. 
Our algorithm is 
the first to incorporate both
approaches, which gives us a broader collection of tools to draw from. 
In addition, we establish a novel connection to \emph{extremal set theory} in the context of graph cut algorithms, which may be of independent interest. Finally, our actual \kcut algorithm resembles the bounded-depth branching algorithms commonly found in \emph{fixed-parameter tractable} algorithms.


Our \kcut algorithm is conceptually simple at a high level. Let
$S^*_1,\lds,S^*_k\s V$ be an arbitrary minimum $k$-cut. We compute a set
$\m A\s 2^V$ of potential subsets of vertices $A\s V$ such that one of
these sets is exactly $S^*_i$ for some $i$. We then branch on each
computed set $A\in\m A$ by guessing $A$ as one component in the targeted
$k$-cut, and then recursively calling \kkcut on $G \setminus A$. It is
clear that this algorithm will return the correct $k$-cut on one of its
branches. Naturally, to obtain an efficient algorithm, we want the size of $\m A$
to be small to ensure a small branching factor, and furthermore, $\m A$
should be computable efficiently. 

\paragraph{A Simpler Case.}
The actual set $\m A$ is complicated
to describe, so to strive for the simplest exposition that highlights
most of our techniques, let us assume (\emph{with} much loss of generality)
that $\m A:=\{ A\s V : w(\pt_GA)\le\f{1.49}kOPT \}$. (Here, any constant
less than $1.5$ will do.) This set $\m A$ works if there exists a
component $S^*_i$ satisfying  $w(\pt_GS^*_i)\le\f{1.49}kOPT$. This may
not always hold, since $\sum_i w(\pt_GS^*_i)=2\,OPT$ and hence we can
only get a bound of $w(\pt_GS^*_i)\le\f2kOPT$ in general---but this is a
simple case considered for the purposes of intuition. It remains to
bound the size of $\calA$, and the time to compute it.

\begin{enumerate}
\item \emph{The Size of $\calA$:} We can bound the size of $\m A$ by
  $O(2^kn)$ using a well-known result in extremal set theory, discussed
  below. But assuming this bound, we pay a branching factor of $O(2^kn)$ to
  decrease $k$ by $1$, which is great, because if we can continue this
  process all the way to $k=0$, then our running time becomes
  $O(2^{k^2}n^{k+O(1)})$. (This bound is much better than the one in
  \Cref{thm:main}, but recall that we obtained it with much loss in
  generality.)
\item \emph{Computing $\m A$:} We run a modified Karger-Stein randomized
  contraction procedure $\poly(n)$ times, and then take the $O(2^kn)$
  sets $A$ with smallest boundary $w(\pt_GA)$. An argument similar to
  Karger and Stein's original analysis shows that our computed set
  contains $\m A$ w.h.p., which is good enough for our purposes.
\end{enumerate}

We now sketch the proof of $|\m A|\le O(2^kn)$, highlighting its
connection to extremal set theory. It will be useful to view each set
$A\in\m A$ also as the ($2$-)cut $(A, V\setminus A)$ in the graph $G$.
\begin{wrapfigure}{R}{0.4\textwidth}
  \includegraphics[width=0.35\textwidth]{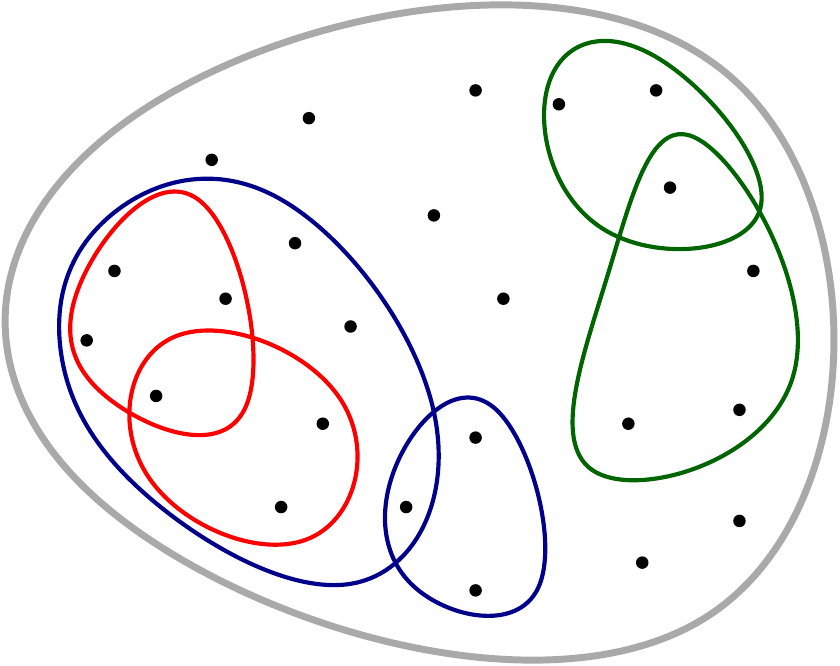}
  \caption{{\small An example with $k=10$; we picked $3$ pairs of
      crossing cuts (i.e., $6$ cuts) each of weight
      $\f{1.49}kOPT$. Because each of them gives $3$ new pieces, we get
      $k = 10$ parts. But the cut edge weight totals $\f{6 \times 1.49}{k}OPT < OPT$, a
      contradiction.}}
  \label{fig:conform}
\end{wrapfigure}
For a contradiction, suppose that $|\m A|>O(2^kn)$. Since $|\m A|>2n$, a
simple result in extremal set theory says that there exist two
such sets $A_1,A_2\in\m A$ that \emph{cross}. Hence, if we cut out the
edges in $\pt_GA_1\cup\pt_GA_2$, we obtain a $4$-cut, not a
$3$-cut. This means that we obtain $3$ new components for the price of
$2 \cd \frac{1.49}kOPT = \f{2.98}kOPT$. This amortizes to a cost of
$\f{2.98}{3k} OPT$ per  additional component. If we can repeat this
process---always finding two such crossing cuts whose removal introduces
$3$ additional components---then we eventually obtain roughly $k$
components for roughly $(1-\e)OPT$, contradicting our choice of
$OPT$. (See Figure~\ref{fig:conform}.) In \S\ref{sec:extremal-bounds}, we prove that this process is possible whenever
$|\m A|>O(2^kn)$, implying the desired extremal bound on $|\m A|$.




\paragraph{The General Case:} Let's try to remove the simplifying
assumption we made. 
What if every component $S^*_i$ satisfies $w(S^*_i) \ge \f{1.5}kOPT$? We
could try to take $\m A:=\{ A\s V : w(\pt_GA)\le\f2kOPT \}$, which would
certainly contain some $S^*_i$. We do take this approach, bounding the
size of $\m A$ (which is substantially more technical), but in this
case, it is possible that $|\m A|=\Om(n^2)$. Branching would result in
an overhead of $\Om(n^2)$, leading to an $\Om(n^{2k-O(1)})$ time
algorithm, which is no good. Our solution is to still branch on each set
in $\m A$, but not start over completely in each recursive step. Rather,
we keep track of a global measure of progress that amortizes our
branching cost over the $k$ recursive calls. In particular, we maintain
a nonnegative \emph{potential function} such that every time we branch
on a large set $\m A$ (and thus pay an expensive branching factor), the
potential function decreases by a lot. This ensures that we do not
branch expensively too often.

Our measure of progress is obtained via the tree packing result of Thorup. 
For a fixed min $k$-cut $\m S^* = \{ S^*_1,\lds,S^*_k \} \s V$, 
we start from a spanning tree $T$ that crosses $\m S^*$ (i.e., connects $S^*_i \neq S^*_j$) at most $2k-2$ times,
and run the previous branching procedure by guessing one component $S^*_i$ at a time. 
Our potential function is based on the current value of $k$ and the number of edges of the current tree crossing $\m S^*$. 
Whenever we branch on an expensive set $\m A$ (say, $|\m A|=\Om(n^2)$), we ensure $\m A$  contains some $S^*_i$ that cuts many edges in $T$, which decreases the potential function substantially. 

\paragraph{Connection to VC-Theory:} 
Consider the set system $(V, \m A)$ with universe $V$ and subsets $\m
A$. The set theory result above (saying that any set system with more
than $2n$ sets has a pair of crossing sets) implies that the {\em dual
  VC dimension} of $(V, \m A)$ is at least $2$ when $| \m A | > 2|V|$;
there exist two sets $A_1, A_2 \in \m A$ such that in their Venn
diagram, all four cells are nonempty. (The dual VC dimension is the
standard VC dimension of the dual set system.)  In
\S~\ref{sec:extremal-results} and \S\ref{sec:extremal-problem}, we prove
an analogous result for the dual VC dimension $3$, which has not been
studied to the best of our knowledge.  We also prove better bounds when
we want three sets $\m A$ {\em partially shattered} (i.e., at least $x$
out of $8$ cells in the Venn diagram are nonempty for some $x < 8$).
This saves the branching cost and allows the global inductive analysis
based on the potential function.

\subsection{Other Related Work}
\label{sec:related-work}

The \kcut problem is NP-hard when $k$ is part of the input~\cite{GH94}.
Karger and Stein gave a randomized Monte-Carlo algorithm with runtime
$O(n^{(2-o(1))k})$, via random edge-contractions~\cite{KS96}. On the
deterministic front, Thorup improved the $O(n^{4k + o(1)})$-time
algorithm of~\cite{KYN06} to $\tilde{O}(n^{2k})$-time, based on tree
packings~\cite{Thorup08}. These approaches also can be used to enumerate
all minimum $k$-cuts, and hence to show that there are at most
$O(n^{2k})$ such cuts. Our work~\cite{GLL18focs} gave a deterministic
algorithm that runs in time roughly $O(k^{O(k)} n^{(2\om/3)k})$ for
bounded integer weights, where $\om \leq 2.3738$ is the
matrix-multiplication constant, but it did not bound the number of
minimum $k$-cuts. Finally, better algorithms are known for small values
of $k \in [2, 6]$~\cite{NI92, HO92, BG97, Karger00, NI00, NKI00,
  Levine00}.  The Karger-Stein algorithm was recently extended to
\textsc{Hypergraph} k-\textsc{Cut}~\cite{ghaffari2017random,
  chandrasekaran2018hypergraph}, which also gave a bound on the
number of minimum $k$-cuts.
For the minimum cut ($k = 2$), the number of and the structure of approximate
min-cuts also have been studied~\cite{henzinger1996number, benczur2008deformable}. 

\paragraph{Approximation algorithms.} The \kcut problem has several
$2(1-1/k)$-approximations due to~\cite{SV95,NR01,RS02}; the more general
Steiner $k$-cut problem also has the same approximation
ratio~\cite{CGN}.  This approximation can be extended to a $(2 -
h/k)$-approximation in time $n^{O(h)}$~\cite{XCY11}. Chekuri et
al.~\cite{chekuri2018lp} studied the LP relaxation of~\cite{NR01} and
gave alternate proofs for both approximation and exact algorithms with
slightly improved guarantees.  A fast $(2 + \eps)$-approximation
algorithm was also recently given by Quanrud~\cite{quanrud2018fast}. On
the hardness front, Manurangsi~\cite{Manurangsi17} showed that for any
$\eps > 0$, it is NP-hard to achieve a $(2 - \eps)$-approximation
 in time $\poly(n,k)$ assuming the Small Set Expansion
Hypothesis.


\paragraph{FPT algorithms.} Kawarabayashi and Thorup gave the first
$f(OPT) \cdot n^{2}$-time algorithm~\cite{KT11} for unweighted
graphs. Chitnis et al.~\cite{Chitnis} used a randomized color-coding
idea to give a better runtime, and to extend the algorithm to weighted
graphs. Here, the FPT algorithm is parameterized by the cardinality of
edges in the optimal \kcut, not by the number of parts $k$. 
Our past
work~\cite{GuptaLL18, GLL18focs} gave a $1.81$-approximation for \kcut
in FPT time $f(k)\cdot \poly(n)$; this has since been simplified and improved to a
$\frac53$-approximation by Kawarabayashi and Lin~\cite{KL18}. 

\section{Preliminaries}
\label{sec:prelim}


\paragraph{Notations.}
Consider a weighted graph $G=(V,E,w)$.
For any vertex set $S$, let $\partial_G S$ 
(or $\partial S$ if $G$ is clear from the context)
denote the edges with exactly one endpoint in $S$.
For a partition $\calS = \{S_1,\lds,S_r\}$ of $V$, 
let $E_G(\calS) := \bigcup_{S_i \in \calS} \, \partial S_i$. 
 For a collection
of edges $F\s E$, let $w(F):=\sum_{e\in F}w(e)$ be the sum of weights of
edges in $F$. 
For a set of vertices $S \subseteq V$, we also define $w(S) := w(\pt S)$. 
In particular, for a \kcut solution $\{S_1,\lds,S_k\}$,
the value of the solution is $w(E_G(S_1,\lds,S_k))$.
Let $\Conn(G)$ be number of connected components of $G$.

\paragraph{Thorup's Tree packing. }

The algorithm starts from the following result of Thorup~\cite{Thorup08}. 

\begin{theorem}[Thorup's Tree Packing]\label{thm:TreePacking}
  Given a graph $G = (V, E, w)$ with $n$ vertices and $m$ edges, we can compute
  a collection $\m T$ of $\tilde O(k^3m)$ trees in time $\tilde O(k^3m^2)$ such that 
  for every minimum $k$-cut $\m S^* = \{ S^*_1, \dots, S^*_k \}$, 
  $\E_{T \in \m T}[ |E_{T}(S^*_1,\lds,S^*_k)| ]\le 2k-2$, where the expectation is taken over 
  a uniform random tree $T \in \m T$.
  In particular, there exists a tree $T\in \m T$ with  $|E_T(S^*_1,\lds,S^*_k)|\le 2k-2$.
\end{theorem}

Given a fixed minimum $k$-cut $\m S^* = \{ S^*_1, \dots, S^*_k \}$, 
we say a tree $T$ is a \Ttree if it crosses $\m S^*$ at most $2k - 2$ times. 
If we choose a \Ttree $T\in \m T$, we get the
following problem: cut some $\le 2k-2$ edges of $T$ and
then merge the connected components into exactly $k$ components
$S_1,\lds,S_k$ so that $E_G(S_1,\lds,S_k)$ is minimized. Thorup's
algorithm accomplishes this task using brute force: try all possible
$O(n^{2k-2})$ ways to cut and merge, and output the best one. This gives
a runtime of $\tO(n^{2k-2}m)$~\cite{Thorup08}. The natural question is: can we do
better than brute-force?

\paragraph{Enumerating Small Cuts via Karger-Stein.}
Our algorithm also uses a subroutine inspired by the Karger-Stein algorithm,
which can generate all $\alpha$-minimum cuts in a graph, (i.e., those
with cut value at most $\alpha$ times the min-cut in the graph) in time
$\tilde{O}(n^{2\alpha})$. We note that a slight
modification of this algorithm can generate all cuts of size at most
$\alpha \cdot \frac{OPT_H}{h}$, where $OPT_H$ is the 
minimum $h$-cut value, in essentially the same run-time. The following lemma
is proved in \S\ref{sec:karger-stein}.

\begin{restatable}{lemma}{KargerStein}
Let $OPT_H$ be the weight of the optimum
minimum $h$-cut in $H = (V,E,w)$, and let $M:=\frac{OPT_H}{h}$.  For any
$\al\ge0$, there are at most $2^h n^{2\al}$ many subsets $A\s V$ with
$w(\pt_H A)\le\al M$. Moreover, we can output (a superset of) all such
subsets in $O(2^h n^{2\al+O(1)})$ time, w.h.p. 
\label{lem:enumerate} 
\end{restatable}


\section{A Refined Bound for Small Cuts}
\label{sec:refined-bound}

Lemma~\ref{lem:enumerate} above says that when $M = \frac{OPT}{k}$, there are at most $2^k \cdot n^{2\alpha}$ cuts of size $\alpha
M$.  The main theorem of this section, Theorem~\ref{thm:ExtremalCuts},
improves this bound by a factor of $n^{\Omega(1)}$ for four different
values of $\alpha = \nicefrac32 - \gamma, \nicefrac53 - \gamma, 2 -
\gamma, \nicefrac73 - \gamma$ for arbitrarily small constant $\gamma >
0$. This is then useful for our algorithm in \S\ref{sec:recursive-algorithm}.

Our proof strategy is the natural one: suppose $\alpha = 2 - \gamma$.
Lemma~\ref{lem:enumerate} gives a bound of $2^k \cdot n^{4 - 2\gamma}$
cuts, whereas Theorem~\ref{thm:ExtremalCuts} below will show a bound of
$O(2^kn^{3-1/4})$ cuts, which is much better. To prove it, we show that
if there are more than $O(2^kn^{3-1/4})$ cuts of size $\alpha M$, there
are $\approx \nicefrac k2$ of these cuts such that their removal yields
$k$ components, witnessing a feasible solution of weight $\nicefrac k2
\cdot \alpha M = (\nicefrac{k(2 - \gamma)}{2k}) \cdot OPT$, which
is strictly less than $OPT$ when $k > \Theta(1/\gamma)$, yielding the
contradiction.

The formal proof of Theorem~\ref{thm:ExtremalCuts} will consist of two
parts.  In \S\ref{sec:extremal-results}, we will prove that if a set system
has a large enough number of sets, there exist three sets whose {\em
  Venn diagram} has many nonempty regions. (If we require all regions to be
nonempty, the definition is equivalent to the {\em dual VC dimension};
see \S\ref{sec:extremal-results} for details.)  This implies that
starting from the whole graph, in each iteration we can remove edges
corresponding to three cuts and increase the number of connected
components by many more than three (as many as seven).  Finally,
in \S\ref{sec:extremal-bounds} we will show that as long as the number of sets is
large also in terms of $k$, we can iterate this process until we obtain a
$k$-cut cheaper than $OPT$, leading to the contradiction.

\subsection{Extremal Set Bounds}
\label{sec:extremal-results}

In this section, 
we show that if we have a ``large'' number of sets, then there
exist some collection of three sets whose Venn diagram contains many
non-empty regions. 
Using the contra-positive, given certain forbidden configurations, the
number of sets (cuts) must be small.  
In this section, we merely state our bounds, deferring the proofs to
\S\ref{sec:extremal-problem}. First, a classical bound: 

\begin{restatable}{claim}{crossSimple}
  Given a set system $(X,\m R)$ on $|X|=n$ elements. If the number of
  sets $|\m R|>2n-2$ then there exists two sets $A, B \in \m R$ that
  cross: i.e., all four of $A \setminus B$, $B \setminus A$, $A \cap B$, and
  $X \setminus (A \cup B)$ are non-empty.
\label{clm:cross-simple}
\end{restatable}

I.e., if there are more than $2n-2$ sets, then there exist a pair of
sets $A, B$, such that all four regions in their Venn diagram are
non-empty. We now show analogous results for intersections of three
sets.  Our main results show that if there are ``many'' sets, then there
exist three sets whose Venn diagram has ``many'' non-empty regions.

\begin{restatable}[7-ot-of-8 Regions]{theorem}{extremalsevenSimple}
  Let $(X,\m R)$ be a set system with $|X|=n$. There exists a constant
  $c > 0$ such that if $|\m R|> cn^{3-1/4}$, then there exist three sets
  whose Venn diagram has $7$ out of $8$ non-empty regions.
  \label{thm:extremal7-simple}
\end{restatable}

\begin{restatable}[All 8 Regions]{theorem}{extremalSimple}
  Let $(X,\m R)$ be a set system with $|X|=n$. There exists a constant
  $c > 0$ such that if $|\m R|> cn^{4-1/4}$, then there exist three sets
  whose Venn diagram has all $8$ regions being non-empty.
\label{thm:extremal-simple}
\end{restatable}

A few comments: firstly, Theorem~\ref{thm:extremal-simple} can be stated
in terms of the \emph{dual VC dimension} (see
Definition~\ref{def:dual-vc}): any set system $\m R$ with
$|\m R| > cn^{4-1/4}$ has dual VC dimension at least $3$.  Secondly, a
polynomial bound of $O(n^8)$ 
also follows from
VC dimension theory, namely, the upper bound $O(n^{2^d})$ on set systems
with dual VC dimension less than $d$. Finally, the bound cannot
be improved below $\Om(n^3)$, since if $\m R$ is all subsets of $X$ of
size $3$, then $|\m R|=\Om(n^3)$ but no three sets in $\m R$ have all
eight regions non-empty. As mentioned earlier, all proofs are in \S\ref{sec:extremal-problem}.



\subsection{Number of Small Cuts}
\label{sec:extremal-bounds}

We can now use the extremal theorems to bound the number of small cuts in a
graph. 
Fix an optimal $k$-cut $\m S^*=\{S_1^*,\lds,S_k^*\}$ where the total
weight of the cut edges is $OPT := w(E(\m S^*))$. Let
$M:=\frac{OPT}{k}$, and define the \emph{normalized weight} of the cut
$\partial_G A$ to be
\begin{gather}
  \wbar(A) := \frac{w(\pt_G A)}{\nicefrac{M}2}. \label{eq:norm-cut}
\end{gather}
This normalization means the $\wbar(E(\m S^*)) = 2k$, and hence the
normalized cut value of an average part of $\m S^*$ is $4$. 
We prove the main result of this section upper bounding the number of small cuts, 
improving Lemma~\ref{lem:enumerate} that gives $2^k \cdot n^{\wbar(A)}$. 
Note that this theorem is non-constructive, 
but these improved bounds the number of small sets (that we compute by a variant of Karger-Stein procedure)
reduces the number of required branchings in our recursive main algorithm in \S\ref{sec:recursive-algorithm},
improving the overall running time. 


\begin{restatable}[Few Small Cuts]{theorem}{FewCuts}
\label{thm:ExtremalCuts}
Fix a small enough positive constant $\g>0$ and assume that
$k> \Omega(1/\g)$.
The graph $G$ has at most:
\BOn
\im[1.] $O(2^kn)$ subsets $A$ with $ \wbar (A) \le 3-\g$.
\im[2.] $O(2^kn^2)$ subsets $A$ with $ \wbar (A) \le 10/3-\g$.
\im[3.] $O(2^kn^{3-1/4})$ subsets $A$ with $ \wbar (A) \le 4-\g$.
\im[4.] $O(2^kn^{4-1/4})$ subsets $A$ with $ \wbar (A) \le 14/3-\g$.
\EOn
\end{restatable}

\BP The proofs for all four statements follows the same outline. For the
sake of a contradiction, we assume that the statement is false. We then
construct a $k$-cut $\m S^\dag$ with $\wbar(\m S^\dag) < 2k$, and
hence get a contradiction.
We prove statement~(1) here and defer the other proofs to
\Cref{app:small-cuts}.  We construct $\m S^\dag$ in two stages. In the
first stage, we use an iterative process to get an $r_0$-cut $\m S_0$
for some $r_0\in[k-2,k]$ such that $\wbar(E(\m S_0))\le r_0\cd(2-\frac23\g)$.  In
a second stage we then augment this to get a $k$-cut. In the following,
let $\m A^1$ be the set of subsets $A\s V$ with $\wbar(A)<3-\g$, with
$| \m A^1| > c2^kn$ subsets for a large enough $c$.

For the first stage, let $\m S$ be the current cut at some point in the
algorithm, $r$ is a lower bound on the current number of components. We
start with a single part $\m S = \{V\}$, and hence $r \gets 1$. In each
iteration, we increase $r$ by some $r'\in\{2,3\}$ and increase
$\wbar(\m S)$ by $\le r'\cd(2-\frac23\g)$. If we can maintain this until
$r\in[k-2,k]$, we have our desired cut $\m S_0$. Each iteration is
simple: while $r<k-2$, if there exists a subset $A\in \m A^1$ that cuts
two or more components in $\m S$, then we cut the edges of $\pt_G A$
inside $\m S$, and increase $r$ by $2$. Note that $\wbar(\m S)$ increases by at
most $3-\g \le 2\cd (2-\frac23\g)$.  (Technically, the number of connected
components can increase by more than $2$, but that only helps us, and it
is still an $(r+2)$-cut.)

Otherwise, every subset $A \in \m A^1$ either does not cut any component
of $\m S$, or cuts exactly one of them. Since we have $r$ components,
there are $\le 2^r$ subsets that do not cut any component. Moreover, for
a given component $S\in\m S$ and a subset
$\emptyset\subsetneq X\subsetneq S$, there are $\le2^{r-1}$ many subsets
$A\s V$ such that $A\cap S=X$ and $A$ does not cut any component in
$\m S-\{S\}$. For each such subset $A\in\m A^1$ that cuts one component
$S\in\m S$ with intersection $A\cap S=X$, add $A$ to a bucket labeled
$(S,X)$---so each bucket has size $\le2^{r-1}$. As long as
$|\m A^1| > 2^r + 2^{r-1}\cd 2n$, there are $>2^{r-1}\cd 2n$ subsets
cutting exactly one component of $\m S$, so there are $>2n$ nonempty
buckets. Consequently, there exists a component $S\in\m S$ for which
there are $>2|S|$ non-empty buckets $(S,X)$. Put another way, the sets
in $A$ intersect $S$ in more then $2n$ distinct ways.  Now we can use
the (easy) extremal result from \Cref{clm:cross} to infer that this set
system cannot be laminar, and there exist two nonempty buckets $(S,X_1)$
and $(S,X_2)$ such that $X_1$ and $X_2$ cross. Taking one subset from
each bucket (call them $A_1,A_2$) and cutting the edges
$\pt_G A_1\cup\pt_G A_2$ inside $\m S$ increases the number of
components by at least $3$. Hence we can increase $r$ by $3$ at the
expense of increasing $\wbar(\m S)$ by at most
$2 \cd (3-\g)= 3\cd(2-\frac23\g)$. Repeating this, we obtain our desired
$r_0$-cut $\m S_0$, for $r_0 \in [k-2,k]$. This ends the first stage. 

In the second stage of the construction, we iteratively augment $\m S_0$
to a $k$-cut. We let $\m S^\dag$ be the current cut, initialized to
$\m S_0$: for $k-r_0$ iterations, we take a subset $A\in\m A^1$ that
cuts at least one component in $\m S^\dag$ and cut the edges of $\pt A$
inside $\m S^\dag$. (Since $|\m A^1| > 2^k$, such a set must exist.)
Thus,
\begin{align*}
\wbar(\m S^\dag) & \tsty \le (r_0-1)\cd(2-\f23\g) + (k-r_0)\cd (3-\g)
       \le k(2-\f23\g) < 2(k-1) < 2k
\end{align*}
the second inequality using that preceding expression is maximized when $r_0 =
k-2$, and that $k>\f3{2\g}$. This $k$-cut $\m S^\dag$
gives us the desired contradiction, and hence proves statement~(1). The
proofs of statements~(2)-(4) follow the same outline, with the
difference lying in the argument about how the sets in $\m A^i$
intersect the components in $\m S$. Indeed, we use the more
sophisticated extremal bounds given by
\Cref{thm:extremal7,thm:extremal} instead of using the naive bound from
\Cref{clm:cross}.
\EP


\newcommand{\wval}{w}
\newcommand{\fval}{\ell}

\section{The \MinKCut Algorithm}
\label{sec:recursive-algorithm}

In this section, we introduce the \MinKCut algorithm that achieves the
bound from \Cref{thm:main}.
The algorithm $\MinKCut(G, k, F, s)$ outputs all the
minimum $k$-cuts of $G$ that can be achieved by cutting $s$ edges of the
given forest $F$ (resulting in $s+\Conn(F)$ connected components of $G$)
and reassembling them back to $k$ connected components. 
To distinguish from $k$ in recursive steps, 
let $k_0$ be the initial value of $k$ such that our final goal is to find minimum $k_0$-cuts.
Since Thorup's
tree packing theorem (Theorem~\ref{thm:TreePacking}) gives a collection
of $n^{O(1)}$ trees such that every for minimum $k_0$-cut in the original
graph $G$ there exists a tree in the collection that intersects the
$k_0$-cut at most $2k_0 - 2$ times, running $\MinKCut(G, k_0, T, 2k_0 - 2)$ for
every $T$ in the family finds every minimum $k_0$-cut. We need the extra
generality to handle the recursive calls: we allow \MinKCut to take in a
forest $F$ (instead of a tree) and a parameter $s$ that may not be
$2k-2$.

An important feature of our algorithm is that it is unchanged under
scaling the weights. Hence for the sake of simplicity of the analysis, 
throughout this section, 
for each call to \MinKCut$(G, k, F, s)$, 
we will assume that the total weight of the optimal $k$-cut is
$w(E(\m S^*)) = 2k$.
This choice ensures that $w(S) = \wbar(S)$ for all $S$, where $\wbar$ was defined in \S\ref{sec:extremal-bounds}.
Finally, let $\g>0$ be a small enough constant throughout this section, whose value will be specified at the end.

\subsection{The \Enum Helper Function}

We use the function \Enum which, given a graph $G$ and parameters
$\beta$ and $N$, outputs all (2-)cuts in $G$ with weight at most
$\beta$ (recall that we normalized weights so that $OPT = 2k$). 
By \Cref{lem:enumerate}, there are at most
$2^k n^{\beta}$ such cuts, and they can be found in
$O(2^k n^{\beta + O(1)})$ time.  In some cases, we have better bounds on
the number of such cuts using the
(nonconstructive)~\Cref{thm:ExtremalCuts}; we pass such an improved
bound to the algorithm via the
parameter $N$, and hence the algorithm returns at most $N$ cuts.

\begin{algorithm}
\mylabel{EnumCuts}{\texttt{EnumCuts}}
\caption{\Enum$(G,\be,N)$}
\small
\textbf{Input}: weighted graph $G=(V,E)$ on $n$ vertices, $N\in \N\cup\{\infty\}$.

\textbf{Output}: W.h.p., the $N$ subsets $A\s V$ satisfying
$w(\pt_GA)\le\be$ that have the smallest values of $w(\pt_GA)$. If
there are not $N$ such subsets $A$, then output (a superset of) all such subsets $A$.

\textbf{Runtime}: $O(2^kn^{\be+O(1)})$
\begin{algorithmic}[1]
\State Run the algorithm of \Cref{lem:enumerate} with $\al:=\be/2$ and let $\m P$ be the output.

\State \Return the $N$ subsets $A\in\m P$ with the smallest $w(\pt_GA)$, or the entire $\m P$ if $|\m P| \leq N$.
\end{algorithmic}
\end{algorithm}

\subsection{The Algorithm Description}
\label{sec:algo-desc}

\begin{algorithm}[t]
\mylabel{MinKCut}{\texttt{MinKCut}}
\caption{\MinKCut$(G,k,F,s)$}
\small
\textbf{Input}: $G=(V,E)$ is an integer-weighted graph on $n$ vertices, $F$ is a forest, and $s+\Conn(F) \ge k$.


\textbf{Output}: (A superset of) valid partitions $\m S\in\m P_{F,s,k}$ (\Cref{def:valid})
with weight equal to the minimum $k$-cut in $G$. 
\begin{algorithmic}[1]

\State $z(k, s) := s-(1.75+\Th(\g))k$

\State $\m P\gets\emptyset$ \Comment{All partitions found will be stored in $\m P$}

\

\If {$k < \Th(1/\g)$}
  \State $\m P\gets$ all minimum $k$-cuts, enumerated in $O(n^{2k})$ time with Karger-Stein, etc.
\ElsIf{$z(k, s) < 0$} \Comment{Brute force}\label{line:if}
  \State $\m P\gets $ all ways to delete $s$ edges in $F$ and merge into $k$ components. \Comment{$O(k^{s+\Conn(F)}n^{s+O(1)})$ time} \label{line:enumerate}
\Else \Comment{Recursive algorithm}


\State $\m A^0 \gets \{ A \s V : |\pt_F A| = 0 \}$ \Comment{$2^{\Conn(F)}$ of them}
\State $\m A^1 \gets \{ A \s V : |\pt_F A| = 1 \}$. \Comment{$2^{\Conn(F)+1} n$ of them}


\



  \State $\m A^2 \gets \Enum(G,3-\g,\Th(2^kn)) \cap \{ A\s V: |\pt_FA| = 2 \}$   \Comment{$\Th(2^kn)$ bound from Thm~\ref{thm:ExtremalCuts}}
  \State $\m A^3 \gets \Enum(G,4-\g,\Th(2^kn^{3-1/4}))\cap\{ A\s V: |\pt_FA| = 3 \}$ \Comment{$\Th(2^kn^{3-1/4})$ from Thm~\ref{thm:ExtremalCuts}}\label{line:7}
  \State $\m A^4 \gets \Enum(G,\nicefrac{14}3-\g,\Th(2^kn^{4-1/4}))\cap\{ A\s V: |\pt_FA| = 4 \}$ \Comment{$\Th(2^kn^{4-1/4})$ from Thm~\ref{thm:ExtremalCuts}}
  \For {$\el \in [5,s]$}
    \State $\m A^\el \gets \Enum(G,\be_\ell,\infty) \cap \{A\s V: |\pt_FA| =\el\}$, where \Comment{$|\m A^\el|=O(2^kn^{\be_\ell})$ by Lem~\ref{lem:enumerate}}\label{line:10}
 \begin{align*}
\tsty \beta_{\ell} := g_{k, s}^{-1}(\ell) \iff \ell = g_{k, s}(\beta_{\ell}) \ \text{ (see (\ref{eq:g}))}
 \end{align*}
  \EndFor
  
  \For {$\el \in [0,s]$, $A\in\m A^\el$}
    \State \textbf{let} $G'\gets G[V-A]$ \textbf{and} $F'\gets F[V-A]$ \label{line:13}
    \State \textbf{let} $s'\gets s-|\pt_FA|$\label{line:15}
    \State Recursively call $\MinKCut(G',k-1,F',s')$ and add its output to $\m P$\label{line:16}
  \EndFor

\EndIf
\State  \Return $\{\m S\in\m P: w(E_G[\m S])=\min_{\m S'\in\m P}w(E_G[\m S'])\}$

\end{algorithmic}
\end{algorithm}

The algorithm \MinKCut is formally given as pseudocode, but let us
explain the main steps. The algorithm is recursive: at some stage, we
are given a current graph $G$ (think of this as the original graph with
some vertices carved away) and a forest $F$ (think of this as $T$ induced
by the current graph). We want to delete $s$ edges from $F$, then combine the
resulting pieces into a $k$-partition of $V(G)$ to obtain an optimal $k$-cut.
The algorithm considers the ways in which a generic part $S_i^*$ from the
optimal cut $\m S^* = \{S_1^*, S_2^*, \ldots, S_k^*\}$ cuts the
forest. 

For $\ell = 0, \dots, s$, the set $\m A^{\ell}$ is supposed to contain
candidates for $S^*_i$ that cut $F$ in $\ell$ edges. The algorithm tries each 
$A \in \m A^{\ell}$ as one of the optimal parts, and recursively run 
\MinKCut$(G[V - A], k - 1, F[V - A], s - \ell)$. 
Therefore if we are able to show that there exists $S^*_i$ and $\ell$ such that 
$S^*_i \in \m A^{\ell}$, we will try $S^*_i$ and call 
\MinKCut$(G[V \setminus S^*_i], k-1, F[V \setminus S^*_i], s - \partial_F(S^*_i))$, 
and since $\m S^* \setminus \{ S^*_i \}$ is an optimal $(k-1)$-cut in $G[V \setminus S^*_i]$,
the recursive algorithm will eventually output $\m S^*$.

By naively setting $\m A^{\ell}$ to be the set of all subsets $A \subseteq V$ that 
cross $F$ exactly $\ell$ times, $|\m A^{\ell}| \leq 2^{O(k)} n^{\ell}$ for all $\ell$.
Since branching on $A \in \m A^{\ell}$ drops $s$ by $\ell$, 
together with the fact that $\log_n |\m A^{\ell}| \leq \ell + o(1)$, 
we can think $s$ as a {\em potential} and 
easily argue that the total running time is $n^{s+O(1)}$. 
(I.e., For fixed $\ell$, we make $|A^{\ell}| \leq n^{\ell + o(1)}$ 
guesses to decrease the potential by $\ell$.)
However, since
$s$ could be $2k-2$ for a \Ttree, we may not win anything.

To get faster running time, 
we further restrict $\m A^{\ell}$ to subsets that induce cuts of small weights. For small $\ell = 2, 3, 4$, 
we restrict to $\m A^{2}, \m A^{3}, \m A^{4}$ to subsets whose cut weight in $G$ is 
at most $3-\gamma, 4-\gamma, \nicefrac{14}{3}-\gamma$ respectively.
The standard Karger-Stein bound (\Cref{lem:enumerate}) only guarantees upper bounds worse than
$O(2^k n^{\ell})$, but our new bounds on the small cuts (\Cref{thm:ExtremalCuts}) bound their numbers by $O(2^k n), O(2^k n^{3-\nicefrac14}),  
O(2^k n^{4-\nicefrac14})$, so $\log_n |\m A^{\ell}| < \ell$ for $\ell = 2, 3, 4$. 
For $\ell \geq 5$, we set $\m A^{\ell}$ more conservatively so that it contains sets of cut weight 
at most $\beta_{\ell} < \ell$, so $\log_n |\m A^{\ell}| < \ell$ from the Karger-Stein bound. 
Our analysis will show that this choice of $\m A^{\ell}$'s
will make sure that $S^*_i$ is contained in one of them for correctness.
Using a different potential function, it will also prove the desired running time.

Before we go further, let's give a convenient definition:
\begin{definition}[Valid Partitions]
  \label{def:valid}
  Given a forest $F = (V,E_F)$ with $\Conn(F)$ connected components, and
  two integers $k, s$ with $s \geq k-\Conn(F)$, a partition
  $\m S=\{S_1,\lds,S_k\}$ of $V$ is \emph{$(F,s,k)$-valid} if it is
  formed by deleting exactly $s$ edges in the forest $F$ and merging the
  resulting $s+\Conn(F)$  connected components of $F$ into $k$
  components. Let $\m P_{F,s,k}$ be all such $(F,s,k)$-valid
  partitions. 
\end{definition}

\subsection{Overview of the Parameter Selection and Analysis}
\label{sec:overview-intuition}

Suppose we are solving $k$-cut on $G$, where we want to delete a total
of $s$ edges in the forest $F$. By our assumption about scaling the
total weight, observe that the target minimum $k$-cut
$\m S^*=\{S^*_1,\lds,S^*_k\}$ satisfies $\sum_i w(\pt_G S_i^*) = 4k$ and
$\sum_i |\partial_F S_i^*| =2s$. 
(This normalization happens at every recursive call.) 
Suppose we have some ``good'' optimal component
$S_i^*$ that we'd like to guess, and then recurse on the graph
$G[V \setminus S_i^*]$. What properties would we like $S_i^*$ to have?
Here are some observations.
\begin{enumerate}
\item For brevity, define $w := w(\partial_G S_i^*)$, and $\ell := |\partial_F S_i^*|$. Then by
  running \Enum with parameter $w$, we can enumerate a set of $O(2^kn^w)$
  subsets $A\s V$ such that one of them equals this targeted
  $S^*_i$. Then, by branching on each of these subsets, we pay a
  multiplicative branching overhead of $O(2^kn^w)$. Moreover, we can use
  the extremal bounds of \Cref{thm:ExtremalCuts} to do even better: if
  $w < 14/3-\g$, then we can enumerate a set much smaller than
  $O(2^kn^w)$. (For example, if $w \le 3-\g$, then our set has size only
  $O(2^kn)$, using \Cref{thm:ExtremalCuts}.)

\item Secondly, we reduce the parameter $s$ in a recursive call by
  $\ell =  |\partial_F S_i^*|$. Intuitively, a set $S_i^*$ with $\ell \gg w$ is a good
  candidate, since we pay a comparatively small branching factor for
  deleting many edges in $F$.

\item So let's keep track of how much we \emph{gain relative to the
    brute-force algorithm}. Namely, if we pay a branching cost of
  $n^\be$ to delete $\ell$ edges (we ignore any constants or $f(k)$ in
  front of the $n^\be$), then since brute-force requires $n^\ell$ time
  to guess the $\ell$ edges in $\pt_FS^*_i$, we are a factor
  $n^{\ell-\be}$ ``ahead'' of brute force. In this case, we measure our
  \emph{gain} as the quantity $\ell - \be$. And we will never branch on
  a set with a negative gain.

\item Finally, when might we \emph{not} be able to make any positive gain?
  Suppose half the $S_i^*$ have $(w=3,\ell=2)$ and the other half have
  $(w=5,\ell=5)$, which means that $s=1.75k$.
  It satisfies our condition $\sum_i w(\pt_G S_i^*) = 4k$ and $\sum_i |\partial_F S_i^*| =2s$,
  but the extremal bounds from
  \Cref{thm:ExtremalCuts} and \Cref{lem:enumerate} only guarantee us
  branching factors of $O(2^kn^2)$ and $O(2^kn^5)$ respectively---in
  both cases there is zero gain. Therefore, if $s\le1.75k$, then in the
  worst case, there may be no choice of set that gives us a positive
  gain, and we might as well do brute-force. Conversely, we will show below
  that as long as $s \ge (1.75 + \Th(\g))k$, positive gain is always
  possible.
\end{enumerate}

\begin{figure}[t]
  \centering
  \includegraphics[width=0.35\textwidth]{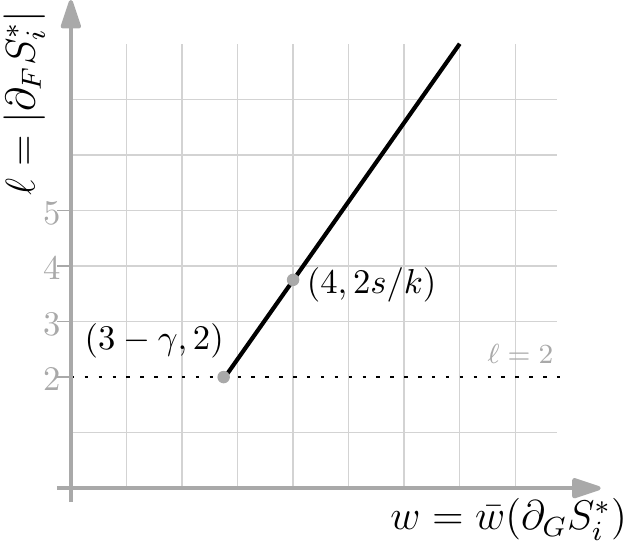}
  \caption{{\small The line $\ell = g_{k, s}(w)$.}}
  \label{fig:2dfigure}
\end{figure}


Now assume $s \ge (1.75 + \Th(\g))k$, and imagine the two-dimensional plane $\R^2$
where we plot the $(\wval,\fval)$ values for the different sets 
$S_i^* \in \m S^*$ as points.
The above discussion suggests a natural strategy of choosing a $S^*_i$, 
which is to draw a line with positive slope and choose $S^*_i$ corresponding to a point above the line. 
To guarantee the existence of a point above the line, we make sure that the lines
pass the centroid of these $k$ points $(4,2s/k)$. 
Two points $(4,2s/k)$ and $(3 - \gamma, 2)$ (instead of $(3, 2)$ for technical reasons)
decide the following line. 
\begin{align}
  \el = g_{k, s}(w) &:= \lp \lp\f{2s}k-2\rp / \lp 1 + \gamma \rp \rp \wval + 8/(1+\gamma) + \f s k \lp 2 - 8 / (1 + \gamma)\rp  \label{eq:g} \\ 
        &= \lp\f{2s}k-2 - O(\gamma) \rp \wval + \lp 8 - \frac{6s}{k} + O(\gamma) \rp.
          \label{eq:line}
\end{align}

See Figure~\ref{fig:2dfigure}. 
The line passes through the centroid $(4, 2s/k)$, so we get the following statement,
whose (easy) proof we omit:

\BCL\label{clm:line}
There exists a set $S^*_i$ such that $\fval \ge g_{k, s}(w)$. 
\ECL

 (We handle the case $\ell < 2$ separately, so only consider the segment from $\ell \geq 2$).
Consider a $S^*_i$ whose $(w, \ell)$ point is above the line guaranteed by \Cref{clm:line}.
Note that the line is above $(5, 5)$ as long as $s \geq (1.75 + \Th(g))k$, and the slope is
$\nicefrac{2s}{k} - 2 - O(\gamma) \geq 1.5 - O(\gamma)$ so for large values of $w, \ell > 5$, 
we have $w < \ell$, and the branching cost $O(n^\beta)$ with $\beta = w$ by \Cref{lem:enumerate} is good.
For small values, $\ell$ can be less than $w$, but our improved bounds 
\Cref{thm:ExtremalCuts} make sure that the branching factor is $O(n^{\beta})$ for $\beta < \ell$.
In any case, our gain $\ell - \beta$ is strictly positive. 

How do we analyze the performance of this algorithm? 
Suppose we run this process from $(k_0,2k_0-2)$ and end up doing brute-force at the point $(k,s)$ with total gain $g$ over the entire process. This means that we have paid a total branching cost of $f(k)n^{(2k_0-2)-s-g}$ so far, leaving a forest $F$ with $\sum_i\fval\le s$. We will then brute-force over the remaining $\le s$ edges for a total running time of $f(k)n^{2k_0-2-g}$. Thus, to obtain the fastest $k_0$-cut possible, it is clear that our objective should be to maximize our total gain over the course of the process (which we can terminate at any point).
Let us define the following quantity that measures how large $s$ is compared to $(1.75 + \Th(\g))k$. 
\begin{gather}
z(k,s) := s-(1.75+\Th(\g))k .\label{eq:z}
\end{gather}
We can view $z$ as a \emph{budget} that starts out as $(0.25-\Th(\g))k_0$ and eventually becomes zero.
We are interested in the amount of gain per unit of budget, which may depend on the current budget.

Recall that whenever $\wval \le 14/3-\g$, our bounds in \Cref{thm:ExtremalCuts} provide better bounds than what is guaranteed in \Cref{lem:enumerate}. We now formalize this below. Define the function
\begin{equation*}
  d(w):=\begin{cases}
        1 & \text{if $w \le 3-\g$},\\
    3-\nicefrac14 & \text{if $3-\g<w\le4-\g$},\\
    4-\nicefrac14 & \text{if $4-\g<w\le\nicefrac{14}3-\g$, and}\\
    w & \text{if $w>\nicefrac{14}3-\g$}.
  \end{cases}
\end{equation*}
By \Cref{thm:ExtremalCuts}, if we branch on $(\wval,\fval)$, then the
gain is $\fval-d(\wval)$ and the difference in budget is
$z(k,s)-z(k-1,s-\fval) = \fval-(1.75+\Theta(\g))$. We now prove the lemma below that
bounds the budget-gain ratio, whose proof is deferred to Appendix~\ref{appendix:4}.

\begin{restatable}{lemma}{gain}
Consider the current state $(k,s)$ with budget
$z=z(k,s)=s-(1.75+\Th(\g))k$, where $s \geq (1.75+\Th(\g))k$, and $k
\leq k_0$. Every
point $(\wval,\fval)$ with $\fval \geq 2$ guaranteed by \Cref{clm:line} satisfies
\[ \f{\fval-d(\wval)}{\fval - (1.75+\Th(\g))} \ge \min \lp \f19,
  \f{4z}{6.5z+4.875k} (1 -  O(\g))\rp \ge \min \lp \f19,
  \f{4z}{6.5z+4.875k_0} (1 -  O(\g)) \rp .\] 
\label{lem:gain} 
\end{restatable}

Note that to handle the guessing of the coordinates $(\wval,\fval)$, the algorithm essentially \emph{guesses all values of $\fval$}. Namely, for each integer value $\fval$, the algorithm enumerates all sets $A\s V$ such that $(w(A),\fval)$ is on or above the line, which is good enough.

\subsection{Correctness}
\label{sec:correctness}

We prove that $\MinKCut(G, k, F, s)$ finds every minimum $k$-cut that
c
an be formed by deleting $s$
edges in $F$ and merging the $s + \Conn(F)$ connected components of $F$
into $k$ components. Fix any such minimum $k$-cut $\m S^*$.  If
$s(k, s) < 0$, then line~\ref{line:enumerate}
enumerates over all possible valid partitions, and hence $\m S^*$ will
be found.  Otherwise, the following lemma shows that at least one
component $S^*_i \in \m S^*$ will be in $\m A^{\ell}$ for some
$\ell \in [0, s]$ so that we can recurse on $G \setminus S^*_i$ and
eventually find $\m S^*$.

\BL
\label{lem:correct}
If the \textbf{\textup{else}} branch in \MinKCut is taken (i.e.,
$z(k, s) \geq 0 \iff s \geq (1.75+\Th(\g))k$), then there exists an
$\el\in[0,s]$ and $A\in\m A^\el$ such that $A\in\m S^*$.  \EL

\BP
If there exists $S^*_i$ whose $\ell$ is $0$ or $1$, $S^*_i \in \m A^{\ell}$. 
When every $S^*_i$ has $\ell \geq 2$, 
by \Cref{clm:line}, there exists $S^*_i$ such that 
$\ell \ge g_{k, s}(w) \iff g^{-1}_{k, s} (\ell) \ge w$.
If $\ell \geq 5$, 
$\beta_{\ell} = g^{-1}_{k, s} (\ell) \geq w$, and
since $\m A^{\ell}$ contains all cuts that crosses $F$ in $\ell$ edges 
and has weight at most $\beta_{\ell}$ in $G$, 
$S^*_i \in \m A^{\ell}$. 
For $\ell = 2, 3, 4$, let $\beta_{2} = 3-\gamma$, $\beta_{3} = 4-\gamma$, $\beta_{4} = 14/3-\gamma$. Since 
$\ell \leq g_{k, s}(\beta_{\ell}) \iff \beta_{\ell} \geq g_{k, s}^{-1}(\ell)$, so again $S^*_i \in \m A^{\ell}$. 
\EP

\subsection{Running Time}
\label{sec:running-time}

We now proceed to the running time analysis. Motivated by \Cref{lem:gain}, we define the following \emph{potential function}:
\begin{equation}
\Phi(k,s) := 
\begin{cases}
\int_{t=0}^{z(k,s)}\min \lp \f19, \f{4t}{6.5t+4.875k_0}(1 - \Th(\g)) \rp dt  & \mbox{if } z(k, s) \geq 0 \\
1 & \mbox{otherwise.}
\end{cases}
\label{eq:pot}
\end{equation}

The function has a discontinuity at $z(k,s)=0$, but this will be convenient later on. Below, we list the technical properties that we need for $\Phi$, whose routine proofs are deferred to Appendix~\ref{appendix:4}.



\begin{restatable}{lemma}{pot}\label{lem:pot}
For values $(k, s)$ such that $z(k, s) \geq 0$, the function $\Phi(k,s)$ satisfies:
\BOn
\im[1.] $\Phi(k,s) \le s$.
\im[2.] $\Phi(k,s) \le \Phi(k,s-1)+1$.
\im[3.] $\Phi(k,s) \le \min\{\Phi(k-1,s), \Phi(k-1,s-1)\}$.
\im[4.] For all $(\wval,\fval)$ satisfying the condition in \Cref{clm:line}, 
\[ \Phi(k,s) \le \Phi(k-1,s-\fval) + \fval-d(\wval). \]
\im[5.] $g_{k,s}\inv(s) \le s - \Phi(k,s) + O(1)$.
\EOn
\end{restatable}

The running time of $\MinKCut$ is bounded by the following recursive
analysis.
 
\BL\label{lem:runtime}
There exist $c_1=O(1)$ and $c_2=O(1/\g)$ such that 
Algorithm $\MinKCut(G,k,F,s)$ takes time \[(c_1\, 2^{k+s+\Conn(F)}))^k \cdot
  n^{s-\Phi(k,s)+c_2}. \] 
\EL

\BP If $k <\Th(1/\g)$, we take the \textbf{if} branch, and Karger-Stein takes time $O(n^{2k})=O(n^{O(1/\g)})$. This meets the bound for $c_2=O(1/\g)$, using that $\Phi(k,s)\le s$ from \Cref{lem:pot}. 

If $s<(1.75 + \Th(\g))k$, we take the \textbf{else if} branch, and the
enumeration on line~\ref{line:enumerate} takes time
\[
k^{s+\Conn(F)} n^{s+O(1)}\le \big(2^{k+s+\Conn(F)}\big)^k \cdot
n^{s-\Phi(k,s)+O(1)}. \]
Here, we use that if $z(k, s) < 0$, then $\Phi(k, s)$ is defined to be $1$. 

We now focus on the case $s \geq (1.75+\Th(\g))k$, applying induction on $k$.
Since all the recursive calls happen for each $A \in \m A^{\ell}$ for some $\ell \in [0, s]$, we examine each $\m A^{\ell}$
and bound the running time of the recursive calls based on $A \in \m A^{\ell}$. 

Recall that $\beta_\ell$ was defined to be $g_{k, s}^{-1}(\ell)$ for $\ell \geq 5$.
Extend this definition so that $\beta_0 = 0, \beta_1 = \beta_2 = 1, \beta_3 = 3-\nicefrac14, \beta_4=4-\nicefrac14$. 
Observe the following:
\BE
\im $|\m A^\el|\le O(2^kn^{\beta_{\el}})$ by \Cref{lem:enumerate} and \Cref{thm:ExtremalCuts}.
\im For every $A\in\m A^\el$ where $\ell \geq 0$, we define $s' = s - |\pt_F A|$
and $F' = F[V - A]$ in lines~\ref{line:13}--\ref{line:15}. This ensures
that $s'+\Conn(F')\le s+\Conn(F)$. 
\EE

The number of recursive calls on line~\ref{line:16} is at most
$|\m A^\el|$, and by induction, the recursive call corresponding to
$A\in\m A^\el$ has runtime
$(c_12^{(k-1)+s'+\Conn(F')})^{k-1}\cd n^{s'-\Phi(k-1,s')+c_2}$. The
total time of these $|\m A^\ell| \leq O(2^kn^{\beta_{\ell}})$ recursive calls for this value is $\el$ is at most
\begin{gather}
 O(2^kn^{\beta_{\ell}}) \cdot (c_12^{k+s+\Conn(F)})^{k-1}\cdot n^{s'-\Phi(k-1,s')+c_2}\label{eq:0}
\end{gather}

Let us focus on the exponent of $n$ in this expression, $\beta_{\ell}+s'-\Phi(k-1,s')+c_2$.
\Cref{lem:pot} shows that $\beta_{\ell}+s'-\Phi(k-1,s') \leq s-\Phi(k, s)$. 

Substituting into~(\ref{eq:0}), we get that for each $\el\in[0,s]$, the recursive calls take total time
\begin{gather}
  O(2^k) \cdot (c_12^{k+s+\Conn(F)})^{k-1} \cdot n^{s-\Phi(k,s)+c_2}. \label{eq:4}
\end{gather}
Now to bound the time for the nonrecursive part of the algorithm. By
the definition of \Enum and \Cref{lem:enumerate}, the runtime for 
lines~\ref{line:7}~to~\ref{line:10} is dominated by the runtime for
$\Enum(G, \beta_s, \infty)$, which is 
\begin{gather}
  n^{\beta_s + O(1)} \leq n^{s-\Phi(k,s)+c_2}, \label{eq:5}
\end{gather}
for large enough constant $c_2$, using item (6) of \Cref{lem:pot}.

Using~(\ref{eq:4}) to bound the time for recursive
calls, and~(\ref{eq:5}) for the non-recursive part, the total time of
\MinKCut is at most 
\begin{gather*}
  s \cdot O(2^k) \cdot (c_12^{k+s+\Conn(F)})^{k-1} \cdot
  n^{s-\Phi(k,s)+c_2} ~~\leq~~ (c_12^{k+s+\Conn(F)})^{k} \cdot n^{s-\Phi(k,s)+c_2},
\end{gather*}
for large enough $c_1$. This completes the induction and finishes the lemma.
\EP

Now using Thorup's tree packing result from \Cref{thm:TreePacking}, we
obtain an instance $(G,k_0,F,s)$ with a single tree and hence
$\Conn(F)=1$, where the optimal solution cuts this tree in at most
$2k_0-2$ edges. We can try all choices of $s$ from $k_0-1$ to $2k_0-2$. By item (2) of \Cref{lem:pot}, the value $s-\Phi(k_0,s)$ is increasing in $s$, so the runtime will be maximum when $s = 2k_0-2$. Our final running time is therefore
\begin{align*}
  2^{O(k_0^2)} \cdot n^{s - \Phi(k_0,2k_0-2)+O(1)}.
\end{align*}

It remains to compute $\Phi(k_0,2k_0-2)$. If we let $f(t) := \min \lp \f19, \f{4t}{6.5t+4.875k_0}(1 - \Th(\gamma)) \rp$, then the two terms inside the $\min(\cd,\cd)$ are equal when $t$ equals $T:=\f{4.875}{29.5(1-\Th(\g))} k_0$. That is, 
$f(t) = 1/9$ for $t \geq \f{4.875}{29.5(1-\Th(\g))} k_0$ and $f(t)=
\f{4t}{6.5t+4.875k_0}(1 - \Th(\gamma))$ otherwise. Integrating $\Phi(k_0,s)$ for $z=z(k_0,s) < k_0T$, we obtain
\[ \Phi(k_0,s) = \int_{t=0}^{z} \f{4t}{6.5t+4.875k_0}(1 - \Th(\gamma))  = \f{(1 - \Th(\gamma))}{1.625} (z/k_0 - 0.75 \ln ( 4z/k_0+ 3 ) + 0.75 \ln 3 ) k_0. \]
Therefore, for a value of $s$ satisfying $z(k_0,s) > k_0T$, we have
\[ \Phi(k_0,s) = \f{(1 - \Th(\gamma))}{1.625} (T - 0.75 \ln ( 4T+ 3 ) + 0.75 \ln 3 ) k_0 + \f19(z - k_0T) .\]
Plugging in $s:=2k_0-2$, we obtain
\begin{align*}
\Phi(k_0,2k_0-2) &= \f1{1.625} \cd \lp\f{4.875}{29.5} - 0.75 \ln\lp4\cd\f{4.875}{29.5} + 3\rp + 0.75 \ln 3\rp k_0 + \f19\lp0.25k _0- \f{4.875}{29.5}k_0\rp - O(\g k_0)
\\&\approx (0.0192055688 - O(\g)) k_0.
\end{align*}

Assuming that $k_0$ is large enough, set $\g>0$ so that
\[s - \Phi(k_0,2k_0-2)+O(1) \approx (2k_0-2) -(0.0192055688 - O(\g))k_0 + O(1)  \le 1.981k_0.\]
Finally, plugging in this bound of $s - \Phi(k_0,2k_0-2)+O(1)$ into \Cref{lem:runtime} proves \Cref{thm:main}. 

\section{The Extremal Problem}
\label{sec:extremal-problem}

In this section we will prove the extremal theorems
(Theorems~\ref{thm:extremal7-simple} and~\ref{thm:extremal-simple}) from
\S\ref{sec:extremal-bounds} about the size of certain set systems that
do not have certain intersection patterns. The results of this section
may be read independently of the other sections, if desired.

Since we will be talking about two kinds of sets, one over a universe
$X$ and another over a small universe $[k]$, we use the vocabulary of
range spaces commonly used in the computational and discrete geometry
literature. A \emph{range space} $(X,\calR)$ is just a set system
over the ground set $X$, where the sets in $\calR$ are called \emph{ranges}.
The following notion of \emph{representation} will be useful to state
our results formally and concisely: please refer to
Figure~\ref{fig:repr} as well.

\BD[Represent] Let $X$ be a universe of elements. Let $k\in\N$ be a
positive integer and $\m S\s 2^{[k]}$ be a set of subsets of $[k]$. For
given ranges $R_1,\lds,R_k\s X$ and subset $S\in\m S$, we say that
$R_1,\lds,R_k$ \defn{witnesses} $(k,S)$ if there exists element $x\in X$
such that for all indices $i\in[k]$, \[i\in S\iff x\in R_i.\] We say
that a $k$-tuple of ranges $(R_1,\lds,R_k)$ \defn{$\al$-witnesses}
$(k,\m S)$ in $X$ if they witness at least an $\al$-fraction of subsets
$S\in\m S$. We drop the ``in $X$'' if the universe $X$ is clear from
context.

Now consider a range space $(X,\m R)$. We say that $\m R$
\defn{$\al$-represents} $(k,\m S)$ in $X$ if there exists subsets
$R_1,\lds,R_k\in\m R$ that $\al$-witness $\m S$.  \ED

\begin{figure}[H]
  \centering
\begin{tikzpicture}[scale=1]
\node at (-2.5,4.2063) {$R_1$};
\node at (-4,1) {$R_2$};
\node at (-0.963,0.9934) {$R_3$};
\tikzstyle{every node} = [circle, fill, scale=.3]
\draw  (-3,2) ellipse (1 and 1);
\draw  (-2,2) ellipse (1 and 1);
\draw  (-2.5,2.8567) ellipse (1 and 1);
\node at (-2.6109,2.4676) {};
\node at (-2.5573,2.1387) {};
\node at (-2.3126,2.3375) {};
\node at (-3.1081,2.6588) {};
\node at (-2.695,3.3702) {};
\node at (-2.3355,3.3626) {};
\node at (-3.8425,3.2631) {};
\node at (-1.5323,1.7562) {};
\end{tikzpicture}
\caption{\emph{\small A Venn diagram representation of three ranges
    $R_1,R_2,R_3$ on a universe $X$ of $8$ elements. The tuple
    $(R_1,R_2,R_3)$ $\nicefrac58$-witnesses $\Three$,
    $\nicefrac47$-witnesses $\Thre$, and $1$-witnesses $(3,\{ \emptyset,
    \{1\}, \{3\}, \{1,2\}, \{1,2,3\} \})$.}}
\label{fig:repr}
\end{figure}

Again, Figure~\ref{fig:repr} may be useful in understanding the notion.
Readers familiar with the notion of VC-dimension will recognize that if
some range space $(X,\m R)$ $1$-represents the set system
$(k, 2^{[k]})$, this corresponds to the \emph{dual range space}
$(\m R, \m R^*)$ having VC dimension at least $k$. The notion of
witnesses and representation we define above is therefore a more refined notion than
the usual notion of (dual) shattering. While we do not need familiarity
with any of these connections, we refer to, e.g.,~\cite[\S5.1]{Mat-book} for
more details on VC dimension and dual shatter functions.

\begin{restatable}{claim}{cross}
Let $(X,\m R)$ be a range space with $|X|=n$. If $|\m R|>2n-2$ then there exist two ranges in $\m R$ that cross.
\label{clm:cross}
\end{restatable}

In this section, we prove Theorems~\ref{thm:extremal7-simple}
and~\ref{thm:extremal-simple} on extremal set bounds. First, let us
restate them in terms of the above notation.

\begin{restatable}[$\nicefrac78$-representation]{theorem}{extremalseven}
  Let $(X,\m R)$ be a range space with $|X|=n$. There exists a positive
  constant $c$ such that if $|\m R|> cn^{3-1/4}$, then $\m R$
  $\nicefrac78$-represents \Three.
\label{thm:extremal7}
\end{restatable}


\begin{restatable}[$1$-representation]{theorem}{extremal}
  Let $(X,\m R)$ be a range space with $|X|=n$. There exists a positive
  constant $c$ such that if $|\m R|> cn^{4-1/4}$, then $\m R$
  $1$-represents \Three.
\label{thm:extremal}
\end{restatable}

We can rephrase Theorem~\ref{thm:extremal} in terms of the concept \emph{dual VC dimension} from VC dimension theory. The following is a simple, equivalent definition of dual VC dimension in terms of our notion of representation:

\BD[Dual VC dimension]
\label{def:dual-vc}
A range space $(X,\m R)$ has \emph{dual VC dimension $d$} if $d$ is the largest integer such that $\m R$ $1$-represents $(d,2^{[d]})$.
\ED

Thus, \ref{thm:extremal} implies the following extremal bound concerning range spaces of dual VC dimension at least $3$:

\BC\label{cor:extremal}
Let $(X,\m R)$ be a range space with $|X|=n$. There exists a positive
  constant $c$ such that if $|\m R|> cn^{4-1/4}$, then $\m R$
  has dual VC dimension at least $3$.
\EC


We first develop some basic tools in \S\ref{sec:warm-up} to give weaker
results; the proofs of the above theorems then appear in
\S\ref{sec:repr-theorem}. Before we proceed, here are some standard
notions and a basic result about the maximum number of sets in a laminar
set system.

\BD[Crossing Sets]
Let $X$ be a universe of elements. We say that two ranges $A,B\s X$ \defn{cross} in $X$ if $(A,B)$ $1$-witnesses $(2,2^{[2]})$ in $X$. That is, there is at least one element in each of the sets $A\bs B$, $B\bs A$, $A\cap B$, and $\ol{A\cup B}$.
\ED
\BD[Cutting Sets]
Let $X$ be a universe of elements. We say that range $A$ \defn{cuts} range $B$ if there is an element in each of the sets $A\cap B$ and $B\bs A$.
\ED
Using these definitions, we can easily prove Claim~\ref{clm:cross} restated below. 

\cross*

\BP
Fix an arbitrary element $x\in X$. Let $\m R'$ be the set $\{R:R\in\m R,x\notin R\} \cup \{\ol{R}:R\in\m R,x\in R\}$, so that $(X\bs\{x\},\m R')$ is a range space. We have $|\m R'|\ge |\m R|/2$, since for every pair of ranges $R,\ol{R}$ satisfying $\{R,\ol{R}\}\cap\m R\ne\emptyset$, one of $R,\ol{R}$ is in $\m R'$. Since $|\m R'|\ge |\m R|/2>n-2=|X\bs\{x\}|-1$, $\m R'$ is not a laminar set of ranges, so there exist $A,B\in\m R'$ such that $A\bs B$, $B\bs A$, and $A\cap B$ are nonempty. Since $x\in\ol{A\cup B}$, the sets $A,B$ cross in $X$. It is easy to see that any of $(A,\ol{B})$, $(\ol{A},B)$, and $(\ol{A},\ol{B})$ also cross in $X$. Since for one of these pairs, both ranges are in $R$, the claim follows.
\EP

\subsection{Warm-up}
\label{sec:warm-up}

In this section, we prove simpler results while developing some basic
machinery. The ideas here are similar to those used, e.g., in proofs of
the Sauer-Shelah theorem about the VC dimension of set systems.

\BL[Extension Lemma]\label{lem:vcd}
Let $(X,\m R)$ be a range space with $|X|=n$. Suppose the following statement is true, for some fixed constant $c$, nonnegative integers $d,k,r$, and subset $\m S\s 2^{[k]}$ with $r < |\m S|$:
\BOn
\im[1.] If $|\m R| > cn^d$, then $\m R$ $(\nicefrac{r}{|\m S|})$-represents $(k,\m S)$.
\EOn
Then there exists constant $c'$ depending on $c,d$ such that the following statement is also true:
\BOn
\im[2.] If $|\m R| > c'n^{d+1}$, then $\m R$ $(\nicefrac{r+1}{|\m S|})$-represents $(k,\m S)$.
\EOn
\EL

\BP
It suffices to prove statement (2) for $n$ large enough: by setting $c'\ge 2^{n_0}$ for some $n_0\ge0$, the statement is true for all $n\le n_0$, since $|\m R|>c'n^{d+1}$ is impossible if $|X|\le n_0$.

We apply induction for large enough $n$. Pick an arbitrary $x\in X$, and let $\Rboth$ be the set of all $R$ satisfying $R\not\ni x$, $R\in\m R$, and $R\cup\{x\}\in\m R$. 

First, suppose that $|\Rboth| > cn^d$. Then, by the assumption in the lemma, there exists a $r/|\m S|$-witness $(R_1,R_2,\ldots,R_k)$ for $(k,\m S)$. If $(R_1,R_2,\ldots,R_k)$ also $(r+1)/|\m S|$-witnesses $(k,\m S)$, then we are done; otherwise, let $S^*\in\m S$ be a subset not witnessed by $(R_1,R_2,\ldots,R_k)$. For each $i\in[k]$, define $R'_i$ to be $R_i$ if $x\notin S^*$, and $R_i\cup\{x\}$ otherwise; note that $R'_i\in\m R$ always. Then, every subset $S\in\m S$ witnessed by $(R_1,R_2,\ldots,R_k)$ is also witnessed by $(R_1',R_2',\ldots,R_k')$, and moreover, $S^*$ is now witnessed by $(R_1',R_2',\ldots,R_k')$. Therefore, $(R_1',R_2',\ldots,R_k')$ $(r+1)/|\m S|$-represents $(k,\m S)$.

Otherwise, suppose that $|\Rboth|\le cn^d$. Define $\Rone$ to be the set of all $R$ satisfying $R\not\ni x$ and exactly one of $R\in\m R$ and $R\cup\{x\}\in\m R$; observe that $|\Rone|+2\cd|\Rboth|=|\m R|$, and that $(X\bs x,\Rone\uplus\Rboth)$ is a range space with $|X\bs x|=n-1$. If there exists a $(r+1)/|\m S|$-witness for $\Rone\uplus\Rboth$, then clearly, this tuple also $(r+1)/|\m S|$-witnesses $\m R$, so assume not. Applying induction on the contrapositive of statement (2) gives $|\Rone\uplus\Rboth|\le c'(n-1)^{d+1}$. Therefore,
\begin{align*}
|\m R|=|\Rone\uplus\Rboth|+|\Rboth| &\le c'(n-1)^{d+1}+cn^d \\&\le c'n^{d+1}-c'\cd Cn^d + cn^d,
\end{align*}
for some constant $C\ge0$ depending on $d$, and assuming that $n$ is large enough. Setting $c'\ge c/C$ gives
\[ |\m R|\le c'n^{d+1}-c'\cd Cn^d+cn^d \le c'n^{d+1}, \]
so the assumption of statement (2) is false, completing the induction.
\EP

\BL\label{lem:worse}
Let $(X,\m R)$ be a range space with $|X|=n$. There exists
constants $c_1,c_2,c_3,c_4$ such that:
\BOn
\im[1.] If $|\m R|> c_1n$, then $\m R$ $\nicefrac58$-represents \Three.
\im[2.] If $|\m R|> c_2n^2$, then $\m R$ $\nicefrac68$-represents \Three.
\im[3.] If $|\m R|> c_3n^3$, then $\m R$ $\nicefrac78$-represents \Three.
\im[4.] If $|\m R|> c_4n^4$, then $\m R$ $1$-represents \Three.
\EOn
\EL
\BP
We first prove (1). Let $c_1:=16$. Since $|\m R|> 16n$, there exists $A,B$ that cross, by Claim~\ref{clm:cross}. If $C$ is any set other than the union of some subset of $\{A\bs B, B\bs A, A\cap B, \ol{A\cup B}\}$, then $(A,B,C)$ is a $\nicefrac58$-witness for \Three. There are at most $2^4=16$ such unions and $|\m R|> 16n\ge 16$, so there is such a choice for $C$.

This serves as the base case: we can now use the Extension Lemma above:
Statement (2) follows from (1) and an application of \Cref{lem:vcd} with
$d:=1$, $k:=3$, $r:=5$, and $\m S:=\three$. Likewise, statement (3)
follows from (2) with $d:=2$ and $r:=6$, and statement (4) follows from
(3) with $d:=3$ and $r:=7$.  \EP

Observe that statement~(4) of \Cref{lem:worse} is weaker than
\Cref{thm:extremal} by an $n^{1/4}$ factor. If we were to stick to the
proof strategy in \Cref{lem:worse}, we would want to prove improved
statements~(1) to (3) in order to get an improved statement~(4).
However, statements~(1) and (2) of \Cref{lem:worse} are tight, and
become false if we replace $|\m R|\ge c_dn^d$ with
$|\m R|\ge c_dn^{d-\e}$ for $d=1,2$ and some $\e>0$. For example, if
$\m R$ is all ranges of size $2$ in $X$, then $\m R$ has size $\bn n2$
but still does not $\nicefrac68$-represent \Three. However, statement (3) does not suffer from a simple, matching lower bound, and indeed, as promised by \Cref{thm:extremal7}, the bound can be improved by $n^{1/4}$. Our main focus will be to prove this improved upper bound on statement (3), from which the improved statement (4) from \Cref{thm:extremal} will follow.

\subsection{Proof of the Extremal Theorems}
\label{sec:repr-theorem}

The main idea behind our improved proofs is prove statements that give a
finer-grained control over the occupied regions of the Venn
diagrams. Since some regions of the Venn diagram are ``easy'' to achieve
(say the intersection of all sets, or the complement of their union), we
seek results that show that with enough sets, there exist three sets
such that many of the ``difficult'' regions are occupied. 

For example, the first structure lemma shows that with linear number of
sets, one can have four Venn diagram regions be occupied. Note that
statement~(1) of \Cref{lem:worse} already showed how to achieve five out
of eight regions. But the lemma below ensures that the four occupied
regions do not include the ``easy'' regions $[3]$ or $\emptyset$;
this makes the proof considerably more technical. (We defer its somewhat
unedifying proof
to \S\ref{sec:def-extremal}.)

\BL[First Structure Lemma]
\label{lem:linear} Let $(X,\m R)$ be a range space with
$|X|=n$. There exists positive constant $c$ such that if $|\m R|> cn$,
then $\m R$ $\nicefrac46$-represents \Thr.
\EL

\BC\label{cor:large_n2}
Let $(X,\m R)$ be a range space with $|X|=n$. There exists positive constant $c$ such that if $|\m R|\ge cn^2$, then $\m R$ $\nicefrac56$-represents \Thr.
\EC
\BP
Start with \Cref{lem:linear} and apply the Extension \Cref{lem:vcd} with
parameters $d:=1$, $k:=3$, $r:=4$, and $\m S:=\thr$.
\EP

The next structure lemma is again in the same vein: the statement is
similar to statement~(2) of \Cref{lem:worse}, and in fact seems
quantitatively worse. (It requires a large number of ranges, and also
that ranges have small size.) But again it does not require the ``easy''
set $\emptyset$ to be represented; we will use this flexibility soon.

\BL[Second Structure Lemma: Small Sizes]
\label{lem:small}
Let $(X,\m R)$ be a range space with $|X|=n$, and let $\e\le1$ be a positive constant. Suppose that every range $R\in\m R$ satisfies $|R|\le n^\e$. Then, there exists positive constant $c$ such that if $|\m R|\ge cn^{2+3\e}$, then $\m R$ $(\nicefrac67)$-represents \Thre.
\EL

\BP The idea of the proof is very natural: since the ranges are small,
we can pick a range $A$ and restrict our attention to the intersections
of other ranges with $A$. If there are many distinct intersections in this
small sub-universe, we inductively get our result. Else there are few
distinct intersections, so on average there are a lot of ranges $R$ that
give the same intersections $R \cap A$. Now we can argue about how the
other ranges intersect $A$ and $\overline{A}$ to prove the result.

For convenience, define $s:=n^\e$. Suppose that $|\m R|\ge cn^{2+3\e}=cn^2s^3$; we will set the constant $c$ later. 
Consider a complete graph $G=(X,E)$ whose vertices are the elements of $X$. We abuse notation, sometimes referring to an edge $(u,v)\in E$ as the two-element set $\{u,v\}\s X$. Call an edge $e=\{u,v\}$ \emph{light} if $e$ is contained in $\le cs^3$ many ranges in $\m R$, and \emph{heavy} otherwise. Also, call a range in $\m R$ \emph{light} if it contains a light edge, and heavy otherwise.

While there exists a light range, remove it from $\m R$, after which some ranges that were previously heavy may become light. We claim that we can remove at most $\bn n2\cd cs^3$ ranges from this iterative operation. For each light range removed, \emph{charge} it to an arbitrary edge inside it that was light when the range was removed. The first time an edge $e$ is charged, it must be light, so $e$ is contained in at most $cs^3$ ranges before it is first charged. Every charge to $e$ reduces the number of ranges containing $e$ by $1$, so edge $e$ is charged at most $cs^3$ times. There are $\bn n2$ edges, leading to at most $\bn n2\cd cs^3$ charges, completing the claim.


Since $|\m R|\ge cn^2s^3>\bn n2\cd cs^3$, there are still ranges left, all of which are heavy; we now work with only these remaining ranges. Let $A$ be a (remaining heavy) range of maximal size.
For each subset $S\s A$, declare a \emph{bucket} labeled with $S$. For each  range $R$ with $R\cap A\ne\emptyset$, add it into the bucket labeled with $R\cap A$. Let $\m R_A\s 2^A$ be the set of labels on nonempty buckets, and let $\m R_A^+\s 2^A$ be the set of labels on buckets of size at least $3$.

First, suppose that $|\m R_A^+| > 2s\ge2(|A|-2)$. Then, there exists nonempty buckets $B_1,B_2\in\m R_A^+$ that cross. Pick a range $R_1\in B_1$ with $R_1\ne B_1$; we can find such a range since bucket $B_1$ has size at least $2$. Pick a range $R_2\in B_2$ with $R_2\ne B_2$ and $R_2\bs A \ne R_1\bs A$; we can find such a range since bucket $B_2$ has size at least $3$. Now consider the tuple $(R_1,R_2,A)$. Since $B_1,B_2$ cross, $(R_1,R_2,A)$ $1$-represents $(3, \{ \{3\}, \{1,3\}, \{2,3\}, \{1,2,3\} \})$. Moreover, by choice of $R_1,R_2$ inside buckets $B_1,B_2$, $(R_1,R_2,A)$ also $2/3$-represents $(3,\{\{1\},\{2\},\{1,2\}\})$. Thus, $(R_1,R_2,A)$ $\nicefrac67$-represents \Thre.

From now on, assume that $|\m R_A^+|\le 2s$. By \Cref{lem:worse} on range space $(A,\m R_A)$, there exists constant $c_3$ such that if $|\m R_A| > c_3\cd |A|^3$, then there exists $(A_1,A_2,A_3)$ that $\nicefrac78$-witnesses \Three in $A$. In particular, if this is the case, then $(A_1,A_2,A_3)$ $\nicefrac67$-witnesses \Thre in $A$. For each $A_i$, let $R_i\in\m R$ be a  range in bucket $A_i$. Then, $(R_1,R_2,R_3)$ $\nicefrac67$-witnesses \Thre in $R$, as desired.

Therefore, we can assume that the number of nonempty buckets, $|\m R_A|$, is at most $c_3\cd |A|^3 \le c_3s^3$.
 Call a bucket \emph{frequent} if it has at  least $ ((c-2c_3)/2)s^2$ ranges inside it.
\BCL
For every $e\in G[A]$, there is a frequent bucket containing $e$.
\ECL
\begin{subproof}
For each edge $e$ in $G[A]$, since $e$ is heavy, it is contained in at least $ cs^3$ many ranges. Since $|\m R_A\bs\m R_A^+|\le|\m R_A|\le c_3s^3$, at most $c_3s^3 \cd 2$ ranges belong to a bucket in $\m R_A \bs \m R_A^+$. This leaves at least $(c-2c_3)s^3$ ranges that belong to buckets in $\m R_A^+$, and since $|\m R_A^+|\le2s$,   there must be a frequent bucket (in $\m R_A^+$) containing $e$. This proves the claim.
\end{subproof}

\begin{figure}
\begin{center}
\begin{tikzpicture}[scale=0.5]

\def \big {(-1,1) ellipse (5 and 5)};
\def \green {(3,7) ellipse (7 and 7)};
\def \cyan {[rotate=20] (-0.3403,1.2719) ellipse (4 and 1)};
\def \red {(-10,6) ellipse (10 and 10)};

    \fill[pattern=north east lines, pattern color=cyan] \cyan;
    \begin{scope}
        \clip \big;
        \fill[pattern=horizontal lines, pattern color=red] \red;
        \fill[pattern=vertical lines, pattern color=green] \green;
        \draw[line width=1,red] \red;
\draw[line width=1,green] \green;
        
    \end{scope}
\draw[line width=1] \big;
\draw[line width=1,cyan] \cyan;

\node at (-5.1428,1.0118) {$B_1$};
\node at (3.2304,1.0717) {$B_2$};
\node at (-.8,-.7) {$B_3$};
\node[above] at (-3.0861,0.2211) {$w$};
\node[above] at (-1.3708,4.2314) {$u$};
\node[above] at (1.7697,1.9847) {$v$};

\tikzstyle{every node} = [circle, fill, scale=.3];

\node at (-3.0861,0.2211) {};
\node at (-1.3708,4.2314) {};
\node at (1.7697,1.9847) {};

\end{tikzpicture}
\caption{The construction of $B_1,B_2,B_3,u,v,w$.}
\label{fig:bucket}
\end{center}
\end{figure}
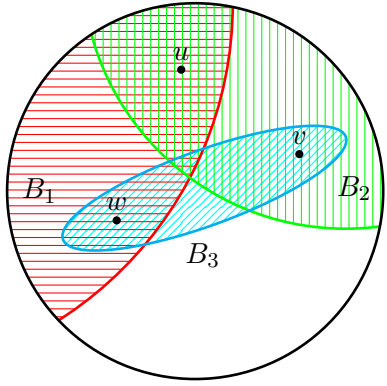

Of all frequent buckets, let $B_1$ be (the label of) the frequent bucket with maximum value of $|B_1|$. We know that $B_1\ne A$, since the bucket labeled $A$ only has one range, namely $A$, so it is not frequent.
Fix two vertices $u\in B_1$ and $v\in A\bs B_1$, and let $B_2$ be a frequent bucket containing edge $(u,v)$; note that $B_2\ne B_1$. Since $B_1$ is the frequent bucket with maximum $|B_1|$, $B_1$ is not contained in $B_2$, so there is a vertex $w$ in $B_1\bs B_2$. Let $B_3$ be a frequent bucket containing edge $(v,w)$. See \Cref{fig:bucket}.

Observe that $(B_1,B_2,B_3)$ $3/4$-witnesses $(3, \{ \{1,3\}, \{2,3\}, \{1,2\}, \{1,2,3\} \} )$ in $A$. Namely, $w\in B_1\cap \ol{B_2}\cap B_3$ and $v\cap \ol{B_1}\cap B_2\cap B_3$ fulfill $\{1,3\}$ and $\{2,3\}$ respectively, and $u\in B_1\cap B_2$ fulfills either $\{1,2\}$ or $\{1,2,3\}$.

\BCL\label{clm:13}
There exist ranges $R_1,R_2,R_3\in\m R$ such that $R_i\cap A=B_i$ for each $i\in[3]$ and $(R_1\bs A,R_2\bs A,R_3\bs A)$ $1$-witnesses $(3,\{ \{1\}, \{2\}, \{3\} \})$ in $X\bs A$.
\ECL

\begin{subproof}
For each $i\in[3]$, pick a random range from the frequent bucket $B_i$. For $i\in[3]$, define event $\Bad_i$ to be the event that $R_i\bs A\s \bigcup_{j\ne i}(R_j\bs A)$. It is clear that if no event $\Bad_i$ holds, then $(R_1\bs A,R_2\bs A,R_3\bs A)$ $1$-witnesses $(3,\{ \{1\}, \{2\}, \{3\} \})$ in $X\bs A$.

To bound $\Pr[\Bad_1]$, fix ranges $R_2,R_3$. Since all ranges $R$ in bucket $B_1$ satisfy $R\cap A=B_1$, the values $R\bs A$ are all distinct. By \Cref{cor:large_n2}, there exists constant $c_2$ such that if at least $c_2|R_2\cup R_3|^2$ many such values $R\bs A$ are contained in $R_2\cup R_3$, then there exist ranges $R_1',R_2',R_3'$ in bucket $B_1$ such that $(R_1'\bs A,R_2'\bs A,R_3'\bs A)$ $\nicefrac56$-represents \Thr in $X\bs A$. Since $R'_1\cap R'_2\cap R'_3 \supseteq B_1\ne\emptyset$,  $(R_1',R_2',R_3')$  $\nicefrac67$-represents \Thre in $X$, proving the lemma. Therefore, we may assume that there are less than $c_2|R_2\cup R_3|^2 \le c_2(2s)^2$ many values $R\bs A$ for range $R$ in bucket $B_1$ that satisfy $R\bs A\s R_2\cup R_3$. The frequent bucket $B_1$ has at least $((c-2c_3)/2)s^2$ ranges, so as long as we have \[((c-2c_3)/2)s^2>3c_2(2s)^2 \iff c>24c_2^2+2c_3,\] we have
\[ \Pr[\Bad_1\text{ given choice of }R_2,R_3] = \Pr[R_1\s R_2\cup R_3\text{ given choice of }R_2,R_3] < \f13. \]
Since $R_2,R_3$ are arbitrary ranges, we have $\Pr[\Bad_1]<1/3$. Repeating the argument for the other two buckets gives $\Pr[\Bad_i]<1/3$ for all $i\in[3]$. Thus, the probability of a bad event is strictly less than $1$, so there is a satisfying choice of $R_1,R_2,R_3$.
\end{subproof}
Thus, the choice of $(R_1,R_2,R_3)$ in Claim~\ref{clm:13} both $3/4$-witnesses $(3, \{ \{1,3\}, \{2,3\}, \{1,2\}, \{1,2,3\} \} )$ in $A$ and $1$-witnesses $(3,\{ \{1\}, \{2\}, \{3\} \})$ in $X \setminus A$, so it is a $\nicefrac67$-witness for \Thre, proving the lemma.
\EP

Recall that the Second Structure \Cref{lem:small} required the sets to
have small sizes. We now give the easy extension to handle all sizes of
sets. 

\BL[Second Structure Lemma: General Sizes]
\label{lem:extremal_n3}
Let $(X,\m R)$ be a range space with $|X|=n$. There exists a positive constant $c$ such that if $|\m R|> cn^{3-1/4}$, then $\m R$ $\nicefrac67$-represents \Thre. 
\EL
\BP
Set $\e:=\nicefrac14$. Call a range in $\m R$ \emph{small} if its size is at most $n^\e$, and \emph{large} otherwise, and let $\m R_\Small$ and $\m R_\Largee$ be the small and large ranges, respectively. If $|\m R_\Small|>(c/2)n^{3-\e}$, then applying \Cref{lem:small} on $\m R_\Small$ proves the lemma, assuming that $c$ is large enough. Otherwise, $|\m R_\Largee|>(c/2)n^{3-\e}$. If so, there is some element $x\in X$ that is in more than $(c/2)n^{3-\e}\cd n^\e/n=(c/2)n^2$ many large ranges. Let $\m R_\Largee^x$ be these large ranges; by \Cref{cor:large_n2}, if $c$ is large enough, then there is tuple $(R_1,R_2,R_3)$ of ranges in $\m R'$ that $\nicefrac56$-witnesses \Thr. Since $x\in R_1\cap R_2\cap R_3$, the tuple $\nicefrac67$-witnesses \Thre, as desired.
\EP

Having proved the structure lemmas, we can turn to proving the main
theorems of this section. We first prove \Cref{thm:extremal7}, restated
below. (This improves on statement~(3) of \Cref{lem:worse}.)

\extremalseven* \BP Set $c:=2c'$, where $c'$ is the constant in
\Cref{lem:extremal_n3}, and suppose that $|\m R|>cn^{3-1/4}$. Following
the proof of Claim~\ref{clm:cross}, fix an arbitrary element $x\in X$,
and let $\m R'$ be the set
$\{R:R\in\m R,x\notin R\} \cup \{\ol{R}:R\in\m R,x\in R\}$, so that
$(X\bs\{x\},\m R')$ is a range space. We have
$|\m R'|\ge |\m R|/2>c'n^{3-1/4}$, since for every pair of ranges
$R,\ol{R}$ satisfying $\{R,\ol{R}\}\cap\m R\ne\emptyset$, one of
$R,\ol{R}$ is in $\m R'$. By the Second Structure \Cref{lem:extremal_n3}, there exists
ranges $R_1,R_2,R_3$ such that $(R_1,R_2,R_3)$ $\nicefrac67$-witnesses \Thre in
$X\bs\{x\}$. Since $x\in\ol{R_1\cup R_2\cup R_3}$, $(R_1,R_2,R_3)$ also
$\nicefrac78$-witnesses \Three in $X$. Finally, one of the eight tuples
obtained by taking or not taking the complement of each $R_i$ gives a
tuple of ranges in $\m R$ that also $\nicefrac78$-witnesses \Three,
proving the theorem.  \EP

An application of the Extension Lemma to the above result proves
\Cref{thm:extremal}, restated below.

\extremal*
\BP
Starting with \Cref{thm:extremal7}, apply the Extension \Cref{lem:vcd}
with parameters $d:=3-1/4$, $k:=3$, $r:=7$, and $\m S:=\three$.
\EP

\subsubsection{Deferred Proofs}
\label{sec:def-extremal}

\begin{figure}\centering
\begin{tikzpicture}[scale=.4]
\node at (-6,14) {$(2)$};

\def \A {(-3,10) ellipse (3 and 3)};
\def \B {(0,10) ellipse (3 and 3)};
\def\C {(-1.5,13) ellipse (2 and 2)};
\draw[blue,line width=1]\C;
\draw[line width=.6]\A;
\draw[line width=.6]\B;
\fill[pattern=horizontal lines, pattern color=cyan]\C;

\node at (-6.5,10) {$A$};
\node at (3.5,10) {$B$};
\node[blue] at (-1.5,15.5) {$C$};

\node[below] at (-5,10) {$a$};
\node[below] at (-1.5,10) {$b$};
\node[below] at (2,10) {$c$};
\node[below] at (-1.5,14) {$d$};

\tikzstyle{every node} = [circle, fill, scale=.3];
\node at (-5,10) {};
\node at (-1.5,10) {};
\node at (2,10) {};
\node at (-1.5,14) {};
\end{tikzpicture}
\begin{tikzpicture}[scale=.4]
\node at (-6,14) {$(3)$};

\node at (0,14) {};
\node at (0,10) {};
\node at (-3,7) {};
\def \C {(0,14) node (v1) {} -- (0,10) -- (0,10) arc (360:270:3) -- plot[smooth, tension=.7] coordinates { (-3,7) (-4.5,9) (-4,14) (-0.9055,15) (v1)}};

\def \A {(-3,10) ellipse (3 and 3)};
\def \B {(0,10) ellipse (3 and 3)};
\draw[blue,line width=1]\C;
\draw[line width=.6]\A;
\draw[line width=.6]\B;
\fill[pattern=horizontal lines, pattern color=cyan]\C;

\node at (-6.5,10) {$A$};
\node at (3.5,10) {$B$};
\node[blue] at (-1.5,15.5) {$C$};

\node[below] at (-5,10) {$c_2$};
\node[below] at (-4,10) {$c_1$};
\node[below] at (2,10) {$b$};
\node[below] at (-1.5,14) {$d$};

\tikzstyle{every node} = [circle, fill, scale=.3];
\node at (-5,10) {};
\node at (-4,10) {};
\node at (2,10) {};

\node at (-1.5,14) {};
\end{tikzpicture}
\begin{tikzpicture}[scale=.4]
\node at (-6,14) {$(5)$};

\node at (0,14) {};
\node at (0,10) {};
\node at (-3,7) {};
\def \C {(0,10) -- (-6,10) -- (-6,10) arc (180:0:3)};

\def \A {(-3,10) ellipse (3 and 3)};
\def \B {(0,10) ellipse (3 and 3)};
\draw[blue,line width=1]\C;
\draw[line width=.6]\A;
\draw[line width=.6]\B;
\fill[pattern=horizontal lines, pattern color=cyan]\C;

\node at (-6.5,10) {$A$};
\node at (3.5,10) {$B$};
\node[blue] at (-3,13.5) {$C$};

\node[below] at (-4,9) {$c_2$};
\node[below] at (-4,11) {$c_1$};
\node[below] at (2,10) {$b$};
\node[below] at (-1.5,9) {$c_3$};

\tikzstyle{every node} = [circle, fill, scale=.3];
\node at (-4,9) {};
\node at (-4,11) {};
\node at (2,10) {};

\node at (-1.5,9) {};
\end{tikzpicture}
\begin{tikzpicture}[scale=.5]
\node at (-6,14) {$(8)$};

\node at (0,14) {};
\node at (0,10) {};
\node at (-3,7) {};
\def \C {(-3 + 3 * 0.8660254 , 10 + 3 / 2) arc (30:300:3) arc (240:210:3) --  (-3 + 3 * 0.8660254 , 10 + 3 / 2)};
\def \CC {(0 - 3 * 0.8660254 , 10 + 3 / 2) arc (150:120:3) arc (60:330:3) -- (0 - 3 * 0.8660254 , 10 + 3 / 2) };

\def \A {(-3,10) ellipse (3 and 3)};
\def \B {(0,10) ellipse (3 and 3)};
\draw[red,line width=1]\C;
\draw[green,line width=1]\CC;
\draw[line width=.6]\A;
\draw[line width=.6]\B;
\fill[pattern=horizontal lines, pattern color=red]\C;
\fill[pattern=vertical lines, pattern color=green]\CC;

\node at (-6.5,10) {$A$};
\node at (3.5,10) {$B$};
\node[red] at (-4,13.5) {$C_1$};
\node[green] at (-2,13.5) {$C_2$};

\node[below] at (-1.5,9.01) {$c_2$};
\node[below] at (-1.5,11.8666) {$c_1$};
\node[below] at (2,10) {$b$};
\node[below] at (-5,10) {$a$};

\tikzstyle{every node} = [circle, fill, scale=.3];
\node at (-1.5,9.01) {};
\node at (-1.5,11.8666) {};
\node at (2,10) {};

\node at (-5,10) {};
\end{tikzpicture}
\begin{tikzpicture}[scale=.5]
\node at (-6,14.2) {$(10)$};

\node at (0,14) {};
\node (v1) at (0,10) {};
\node at (-3,7) {};
\def \C {(-2.8191,11.0261) arc (159.9994:240:3) -- (-1.5,7.4019) arc (-60.0002:-300:3) -- (-1.5,12.5981) arc (119.9998:-120:3) -- (-1.5,7.4019) arc (-60.0002:-20:3) -- (-2.8191,11.0261)};
\def \CC {(-0.1809,11.0261) arc (20.0006:-60:3) -- (-1.5,7.4019) arc (-119.9998:120:3) -- (-1.5,12.5981) arc (60.0002:300:3) -- (-1.5,7.4019) arc (-119.9998:-160:3) -- (-0.1809,11.0261)};
\def \CCC {(-1.5,7.4019) arc (-119.9998:-240:3) -- plot[smooth, tension=.9] coordinates {(-1.5,12.5981)  (-2.1142,13.973) (-0.8106,13.9467) (-1.5,12.5981) } -- (-1.5,12.5981)  {} arc (60.0002:-60:3)};

\def \A {(-3,10) ellipse (3 and 3)};
\def \B {(0,10) ellipse (3 and 3)};
\draw[green,line width=1]\C;
\draw[red,line width=1]\CC;
\draw[blue,line width=1]\CCC;
\draw[line width=.6]\A;
\draw[line width=.6]\B;
\fill[pattern=vertical lines, pattern color=green]\C;
\fill[pattern=horizontal lines, pattern color=red]\CC;
\fill[pattern=north east lines, pattern color=cyan]\CCC;

\node at (-6.5,10) {$A$};
\node at (3.5,10) {$B$};
\node[red] at (-5,13) {$C_1$};
\node[green] at (1,13.5) {$C_2$};
\node[blue] at (-3,14) {$C'$};

\node[below] at (-.5,10) {$c_2$};
\node[below] at (-2.5,10) {$c_1$};
\node[below] at (2,10) {$b$};
\node[below] at (-5,10) {$a$};
\node[above] at (-1.5,12.8) {$d$};

\tikzstyle{every node} = [circle, fill, scale=.3];
\node at (-.5,10) {};
\node at (-2.5,10) {};
\node at (2,10) {};
\node at (-1.5,13.8) {};

\node at (-5,10) {};
\node at (-0.4608,8.4369) {};
\draw (-2.5,7.866) ;

\end{tikzpicture}
\caption{Illustration of some of the cases for the proof of \Cref{lem:linear}.}
\label{fig:deferred}
\end{figure}
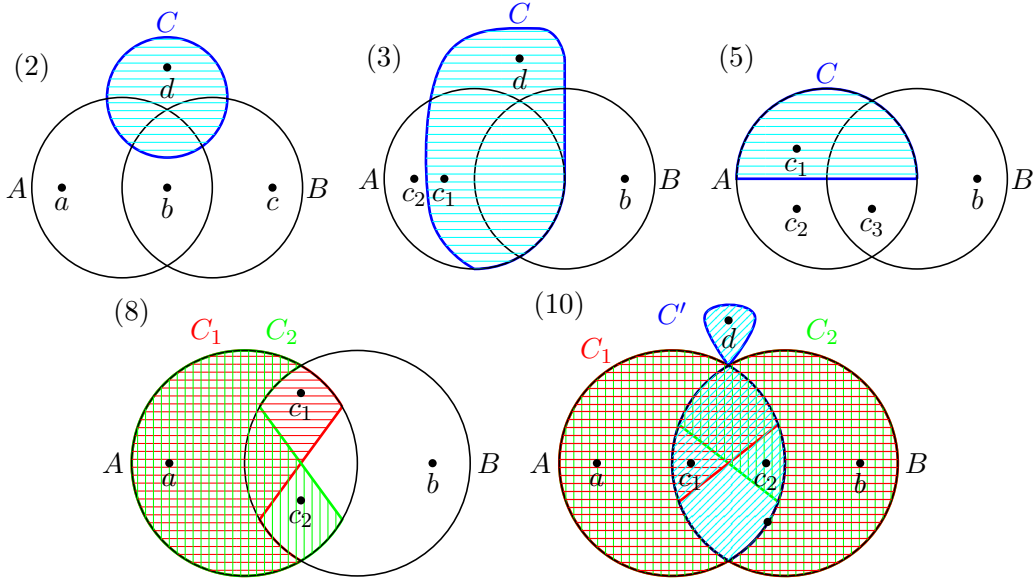

\BP[Proof of \Cref{lem:linear}]
The proof has many cases; see \Cref{fig:deferred} for a visualization of some of the cases.

We induct on $n$, with the case $n=1$ being trivial as long as $c \ge2$, since $|\m R|>2n$ is impossible if $|X|=1$.

Assuming that $c \ge2$, we have $|R|>2n$, so there exist ranges $A,B\in\m R$ that cross. Let $A,B$ be the two crossing sets with minimum $|A\cap B|$. Fix arbitrary elements $a\in A\bs B$, $b\in B\bs A$, 
and $d\in \ol{A\cup B}$.
\BE
\im If no range $R\in\m R$ contains $d$, then remove $d$ from $X$ and apply induction. More formally, set $X':=X\bs\{d\}$; since $(X',\m R)$ is still a range space, we have $|\m R|\le c|X'|$ by induction, so $|\m R|\le c|X|$.
\im If some range $C\ni d$ satisfies $C\not\supseteq A\cap B$, then the set $(A,B,C)$ works, i.e., it $\nicefrac46$-represents \Thr. This is because $a$ fulfills either $\{1\}$ or $\{1,3\}$, $b$ fulfills either $\{2\}$ or $\{2,3\}$, $d$ fulfills $\{3\}$, and there is some element $c'\in (A\cap B)\bs C$ that fulfills $\{1,2\}$.
\im If some range $C\ni d$  cuts $A\bs B$, then the set $(A,B,C)$ works, because $b$ fulfills either $\{2\}$ or $\{2,3\}$, $d$ fulfills $\{3\}$, and there are elements $c_1\in (A\bs B)\cap C$ and $c_2\in(A\bs B)\bs C$ that fulfill $\{1,3\}$ and $\{1\}$, respectively.
\im If some range $C\ni d$ cuts $B\bs A$, then the set $(A,B,C)$ works by a symmetric argument.
\EE

If any element $d\in\ol{A\cup B}$ satisfies (2), (3), or (4), then we are done, so we may assume that none of them do. This implies the following assumption:

\begin{assumption}\label{as:9}
For any range $C\not\s A\cup B$, the intersection $C\cap(A\cup B)$ is one of $A$, $B$, $A\cap B$, and $A\cup B$. 
\end{assumption}

We now focus on ranges $C\s A\cup B$.

\BE\setcounter{enumi}{4}
\im If some range $C\s A\cup B$ cuts $A\bs B$ and satisfies $C\not\supseteq A\cap B$, then the set $(A,B,C)$ works, because $b$ fulfills either $\{2\}$ or $\{2,3\}$, and there are elements $c_1\in (A\bs B)\cap C$, $c_2\in (A\bs B)\bs C$, and $c_3\in (A\cap B)\bs C$ fulfilling $\{1,3\}$, $\{1\}$, and $\{1,2\}$, respectively.
\im If some range $C\s A\cup B$ cuts $B\bs A$ and satisfies $C\not\supseteq A\cap B$, then the argument is symmetric.
\EE
If any range $C\s A\cup B$ satisfies (5) or (6), then we are done, so we may assume that none of them do. This implies the following assumption:

\begin{assumption}\label{as:10}
Every range $C\s A\cup B$ either contains $A\cap B$, or the intersection $C\cap (A\triangle B)$ is one of $\emptyset$, $A\bs B$, $B\bs A$, and $A\triangle B$.
\end{assumption}

We first treat the latter case in Assumption~\ref{as:10} below.
\BE\setcounter{enumi}{6}
\im There are at most $2(|A\cap B|-2)$ ranges $C\s A\cup B$ satisfying $C\cap (A\triangle B)=\emptyset\iff C\s A\cap B$. Otherwise, by Claim~\ref{clm:cross}, there exist two crossing $C_1,C_2$ in $A\cap B$, contradicting the assumption that $A$ and $B$ are the two crossing sets with minimum $|A\cap B|$.
\im If there are more than $2(|A\cap B|-2)$ ranges $C\s A\cup B$ satisfying $C\cap (A\triangle B)=A\bs B\iff A\bs B\s C\s A$, then by Claim~\ref{clm:cross}, there exist two such $C_1,C_2$ that cross. Then, $(C_1,C_2,B)$ works, since $a$ fulfills $\{1,2\}$, $b$ fulfills $\{3\}$, and there are elements $c_1\in C_1\bs C_2$ and $c_2\in C_2\bs C_1$ which are inside $A\cap B$, and which fulfill $\{1,3\}$ and $\{2,3\}$, respectively.
\im If there are more than $2(|A\cap B|-2)$ ranges $C\s A\cup B$ satisfying $C\cap (A\triangle B)=B\bs A\iff B\bs A\s C\s B$, then the argument is symmetric to case (8).
\im If there are more than $2(|A\cap B|-2)$ ranges $C\s A\cup B$ satisfying $C\cap (A\triangle B)=A\triangle B\iff C\supseteq A\triangle B$, then we apply Claim~\ref{clm:cross} for these ranges $C$, obtaining $C_1,C_2$ that cross inside $A\cap B$.
By Assumption~\ref{as:9}, any range $C'\not\s A\cup B$ satisfies $C'\cap (A\cup B)\in \{A,B,A\cap B,A\cup B\}$. First, suppose there exists such a range $C'$ such that $C'\cap (A\cup B) \ne A\cup B$. Then, $(C_1,C_2,C')$ works, because some element $d\in C\bs (A\cup B)$ fulfills $\{3\}$, there are elements $c_1\in C_1\bs C_2$ and $c_2\in C_2\bs C_1$ fulfilling $\{1,3\}$ and $\{2,3\}$, respectively, and one of $a,b$ fulfills $\{1,2\}$, since $C'$ cannot contain both $a$ and $b$.

Otherwise, every range $C'\not\s A\cup B$ satisfies $C'\cap (A\cup B)=A\cup B$. Let $\m R_\Out:=\{C'\bs(A\cup B):C'\in\m R,\,C'\not\s A\cup B\}$ and $\m R_\In:=\{C:C\in\m R,\, C\s A\cup B\}$. Since every $C'\not\s A\cup B$ has the same value of $C'\cap (A\cup B)$, the sets $C'\bs (A\cup B)$ are distinct; in particular, $|\m R_\Out|=|\m R|-|\m R_\In|$.
If either $\m R_\Out$ or $\m R_\In$ $\nicefrac46$-represents \Thr in $\ol{A\cup B}$ or $A\cup B$, respectively, then so does $\m R$ and we are done, so assume otherwise. By induction on the contrapositive statement for $(\ol{A\cup B},\m R_\Out)$ and $(A\cup B,\m R_\In)$, we have $|\m R_\Out|\le c\cd|\ol{A\cup B}|$ and $\m |\m R_\In|\le c\cd|A\cup B|$. Therefore,
\[ |\m R|=|\m R_\Out|+|\m R_\In| \le c\cd |\ol{A\cup B}| + c\cd|A\cup B|=cn ,\] so the assumption of \Cref{lem:linear} is false, completing the induction.
\EE
If any of cases (7) to (10) holds, then we are done, so assume otherwise.
This means that there are at most $8(|A\cap B|-2)$ many ranges $C\s A\cup B$, whose intersection $C\cap(A\cup B)$ is one of $A$, $B$, $A\cap B$, and $A\cup B$. By Assumption~\ref{as:10}, all remaining ranges $C\s A\cup B$ must contain $A\cap B$. In addition, by Assumption~\ref{as:9}, any range $C\not\s A\cup B$ has intersection $C\cap (A\cup B)$ in one of $A$, $B$, $A\cap B$, and $A\cup B$, so in particular, $C$ also contains $A\cap B$. Therefore, there are $\ge |\m R|-8(|A\cap B|-2)$ many ranges in $\m R$ that contain $A\cap B$; define $\m R_\Cont:=\{C\bs(A\cap B):C\in\m R,\, C\supseteq A\cap B\}$ to be these ranges with $A\cap B$ removed. If $\m R_\Cont$ $\nicefrac46$-represents \Thr, then so does $\m R$ and we are done, so assume otherwise. By induction on the contrapositive statement for $(\ol{A\cap B},\m R_\Cont)$, we have $|\m R_\Cont|\le c\cd|\ol{A\cap B}|$. Therefore,
\[ |\m R| \le 8(|A\cap B|-2) + |\m R_\Cont| \le 8|A\cap B| + c\cd |\ol{A\cap B}| \le cn \]
as long as $c\ge8$, so the assumption of \Cref{lem:linear} is false, completing the induction. Thus, setting $c:=8$ concludes the lemma.
\EP



{\small \bibliographystyle{alpha}
\bibliography{refs}}

\newcommand{\etalchar}[1]{$^{#1}$}
\begin{thebibliography}{CCH{\etalchar{+}}16}

\bibitem[ABW15]{abboud2015if}
Amir Abboud, Arturs Backurs, and Virginia~Vassilevska Williams.
\newblock If the current clique algorithms are optimal, so is {V}aliant's
  parser.
\newblock In {\em Foundations of Computer Science (FOCS), 2015 IEEE 56th Annual
  Symposium on}, pages 98--117. IEEE, 2015.

\bibitem[AWW14]{KClique}
Amir Abboud, Virginia~Vassilevska Williams, and Oren Weimann.
\newblock Consequences of faster alignment of sequences.
\newblock In {\em International Colloquium on Automata, Languages, and
  Programming}, pages 39--51. Springer, 2014.

\bibitem[BG97]{BG97}
Michel Burlet and Olivier Goldschmidt.
\newblock A new and improved algorithm for the {$3$}-cut problem.
\newblock {\em Oper. Res. Lett.}, 21(5):225--227, 1997.

\bibitem[BG08]{benczur2008deformable}
Andr{\'a}s~A Bencz{\'u}r and Michel~X. Goemans.
\newblock Deformable polygon representation and near-mincuts.
\newblock In {\em Building Bridges}, pages 103--135. Springer, 2008.

\bibitem[CCH{\etalchar{+}}16]{Chitnis}
Rajesh Chitnis, Marek Cygan, MohammadTaghi Hajiaghayi, Marcin Pilipczuk, and
  Micha{\l} Pilipczuk.
\newblock Designing {FPT} algorithms for cut problems using randomized
  contractions.
\newblock {\em SIAM J. Comput.}, 45(4):1171--1229, 2016.

\bibitem[CGN06]{CGN}
Chandra Chekuri, Sudipto Guha, and Joseph Naor.
\newblock The {S}teiner {$k$}-cut problem.
\newblock {\em SIAM J. Discrete Math.}, 20(1):261--271, 2006.

\bibitem[CQX18]{chekuri2018lp}
Chandra Chekuri, Kent Quanrud, and Chao Xu.
\newblock Lp relaxation and tree packing for minimum $ k $-cuts.
\newblock {\em arXiv preprint arXiv:1808.05765}, 2018.

\bibitem[CXY18]{chandrasekaran2018hypergraph}
Karthekeyan Chandrasekaran, Chao Xu, and Xilin Yu.
\newblock Hypergraph k-cut in randomized polynomial time.
\newblock In {\em Proceedings of the Twenty-Ninth Annual ACM-SIAM Symposium on
  Discrete Algorithms}, pages 1426--1438. Society for Industrial and Applied
  Mathematics, 2018.

\bibitem[GH94]{GH94}
Olivier Goldschmidt and Dorit~S. Hochbaum.
\newblock A polynomial algorithm for the {$k$}-cut problem for fixed {$k$}.
\newblock {\em Math. Oper. Res.}, 19(1):24--37, 1994.

\bibitem[GKP17]{ghaffari2017random}
Mohsen Ghaffari, David~R Karger, and Debmalya Panigrahi.
\newblock Random contractions and sampling for hypergraph and hedge
  connectivity.
\newblock In {\em Proceedings of the Twenty-Eighth Annual ACM-SIAM Symposium on
  Discrete Algorithms}, pages 1101--1114. SIAM, 2017.

\bibitem[GLL18a]{GLL18focs}
Anupam Gupta, Euiwoong Lee, and Jason Li.
\newblock Faster exact and approximate algorithms for $k$-cut.
\newblock In {\em Foundations of Computer Science (FOCS), 2018 IEEE 59th Annual
  Symposium on}, 2018.

\bibitem[GLL18b]{GuptaLL18}
Anupam Gupta, Euiwoong Lee, and Jason Li.
\newblock An {FPT} algorithm beating 2-approximation for \emph{k}-cut.
\newblock In {\em Proceedings of the Twenty-Ninth Annual {ACM-SIAM} Symposium
  on Discrete Algorithms, {SODA} 2018, New Orleans, LA, USA, January 7-10,
  2018}, pages 2821--2837, 2018.

\bibitem[HO94]{HO92}
Jianxiu Hao and James~B. Orlin.
\newblock A faster algorithm for finding the minimum cut in a directed graph.
\newblock {\em J. Algorithms}, 17(3):424--446, 1994.
\newblock Third Annual ACM-SIAM Symposium on Discrete Algorithms (Orlando, FL,
  1992).

\bibitem[HW96]{henzinger1996number}
Monika Henzinger and David~P. Williamson.
\newblock On the number of small cuts in a graph.
\newblock {\em Information Processing Letters}, 59(1):41--44, 1996.

\bibitem[Kar00]{Karger00}
David~R. Karger.
\newblock Minimum cuts in near-linear time.
\newblock {\em J. ACM}, 47(1):46--76, 2000.

\bibitem[KL18]{KL18}
Ken-ichi Kawarabayashi and Bingkai Lin.
\newblock A nearly $5/3$-approximation {FPT} algorithm for min-k-cut.
\newblock {\em Manuscript}, 2018.

\bibitem[Knu73]{knuth2011art}
Donald~E Knuth.
\newblock {\em The Art of Computer Programming, Volume 1: Fundamental
  Algorithms}.
\newblock Addison-Wesley Publishing Company, 1973.

\bibitem[KS96]{KS96}
David~R. Karger and Clifford Stein.
\newblock A new approach to the minimum cut problem.
\newblock {\em Journal of the ACM (JACM)}, 43(4):601--640, 1996.

\bibitem[KT11]{KT11}
Ken-ichi Kawarabayashi and Mikkel Thorup.
\newblock The minimum $k$-way cut of bounded size is fixed-parameter tractable.
\newblock In {\em Foundations of Computer Science (FOCS), 2011 IEEE 52nd Annual
  Symposium on}, pages 160--169. IEEE, 2011.

\bibitem[KYN07]{KYN06}
Yoko Kamidoi, Noriyoshi Yoshida, and Hiroshi Nagamochi.
\newblock A deterministic algorithm for finding all minimum {$k$}-way cuts.
\newblock {\em SIAM J. Comput.}, 36(5):1329--1341, 2006/07.

\bibitem[Lev00]{Levine00}
Matthew~S Levine.
\newblock Fast randomized algorithms for computing minimum $\{$3, 4, 5,
  6$\}$-way cuts.
\newblock In {\em Proceedings of the eleventh annual ACM-SIAM symposium on
  Discrete algorithms}, pages 735--742. Society for Industrial and Applied
  Mathematics, 2000.

\bibitem[LG14]{le2014powers}
Fran{\c{c}}ois Le~Gall.
\newblock Powers of tensors and fast matrix multiplication.
\newblock In {\em Proceedings of the 39th international symposium on symbolic
  and algebraic computation}, pages 296--303. ACM, 2014.

\bibitem[Man17]{Manurangsi17}
Pasin Manurangsi.
\newblock {Inapproximability of Maximum Edge Biclique, Maximum Balanced
  Biclique and Minimum $k$-Cut from the Small Set Expansion Hypothesis}.
\newblock In {\em 44th International Colloquium on Automata, Languages, and
  Programming (ICALP 2017)}, volume~80 of {\em Leibniz International
  Proceedings in Informatics (LIPIcs)}, pages 79:1--79:14, 2017.

\bibitem[Mat99]{Mat-book}
Ji\v{r}\'{\i} Matou\v{s}ek.
\newblock {\em Geometric discrepancy}, volume~18 of {\em Algorithms and
  Combinatorics}.
\newblock Springer-Verlag, Berlin, 1999.
\newblock An illustrated guide.

\bibitem[NI92]{NI92}
Hiroshi Nagamochi and Toshihide Ibaraki.
\newblock Computing edge-connectivity in multigraphs and capacitated graphs.
\newblock {\em SIAM J. Discrete Math.}, 5(1):54--66, 1992.

\bibitem[NI00]{NI00}
Hiroshi Nagamochi and Toshihide Ibaraki.
\newblock A fast algorithm for computing minimum 3-way and 4-way cuts.
\newblock {\em Math. Program.}, 88(3, Ser. A):507--520, 2000.

\bibitem[NKI00]{NKI00}
Hiroshi Nagamochi, Shigeki Katayama, and Toshihide Ibaraki.
\newblock A faster algorithm for computing minimum 5-way and 6-way cuts in
  graphs.
\newblock {\em J. Comb. Optim.}, 4(2):151--169, 2000.

\bibitem[NR01]{NR01}
Joseph Naor and Yuval Rabani.
\newblock Tree packing and approximating {$k$}-cuts.
\newblock In {\em Proceedings of the {T}welfth {A}nnual {ACM}-{SIAM}
  {S}ymposium on {D}iscrete {A}lgorithms ({W}ashington, {DC}, 2001)}, pages
  26--27. SIAM, Philadelphia, PA, 2001.

\bibitem[Qua18]{quanrud2018fast}
Kent Quanrud.
\newblock Fast and deterministic approximations for $ k $-cut.
\newblock {\em arXiv preprint arXiv:1807.07143}, 2018.

\bibitem[RS08]{RS02}
R.~Ravi and Amitabh Sinha.
\newblock Approximating {$k$}-cuts using network strength as a {L}agrangean
  relaxation.
\newblock {\em European J. Oper. Res.}, 186(1):77--90, 2008.

\bibitem[SV95]{SV95}
Huzur Saran and Vijay~V. Vazirani.
\newblock Finding $k$-cuts within twice the optimal.
\newblock {\em SIAM Journal on Computing}, 24(1):101--108, 1995.

\bibitem[Tho08]{Thorup08}
Mikkel Thorup.
\newblock Minimum $k$-way cuts via deterministic greedy tree packing.
\newblock In {\em Proceedings of the fortieth annual ACM symposium on Theory of
  computing}, pages 159--166. ACM, 2008.

\bibitem[Wil12]{williams2012multiplying}
Virginia~Vassilevska Williams.
\newblock Multiplying matrices faster than {Coppersmith}--{Winograd}.
\newblock In {\em Proceedings of the forty-fourth annual ACM symposium on
  Theory of computing}, pages 887--898. ACM, 2012.

\bibitem[WW10]{williams2010subcubic}
Virginia~Vassilevska Williams and Ryan Williams.
\newblock Subcubic equivalences between path, matrix and triangle problems.
\newblock In {\em Foundations of Computer Science (FOCS), 2010 51st Annual IEEE
  Symposium on}, pages 645--654. IEEE, 2010.

\bibitem[XCY11]{XCY11}
Mingyu Xiao, Leizhen Cai, and Andrew Chi-Chih Yao.
\newblock Tight approximation ratio of a general greedy splitting algorithm for
  the minimum $k$-way cut problem.
\newblock {\em Algorithmica}, 59(4):510--520, 2011.

\end{thebibliography}

\appendix

\section{Deferred Proofs}
\label{sec:proofs}

\subsection{Proof of Lemma~\ref{lem:enumerate}}
\label{sec:karger-stein}

Following Karger-Stein's arguments, we prove Lemma~\ref{lem:enumerate} restated below. 

\KargerStein*

\BP The algorithm is simple: start with the original graph, and while
there are more than $h$ vertices left, contract a random edge selected
proportional to its weight. When there are $h$ vertices remaining,
consider all $2^h$ possible subsets of these $h$ vertices. For each
subset $A'$, map it back to a subset $A$ in the original graph (by
taking all vertices in $V$ that were contracted into a vertex of
$A'$), and if $\pt_H A\le\al M$, then add it to our collection of
cuts. Repeat this process $O(n^{2\al}\log n)$ times.

To see correctness, consider an iteration with more than $h$ vertices
remaining, and let $H'=(V',E')$ be the remaining graph. Considering the
$h - 1$ vertices with smallest degree, and the rest of $V'$, when
un-contracted, gives us a $h$-cut of weight, whose weight must be at
least $OPT_H$. Therefore, if $\ol{d}$ is the average degree of $G'$ and
$\ol{d_{h - 1}}$ is the average degree of the $(h - 1)$ smallest-degree
vertices in $G'$, then
\begin{gather}
  \f{OPT_H}{h} \leq  \f{OPT_H}{h - 1} \le \ol{d_{h - 1}} \le \ol{d} =
  \f{2w(E')}{|V'|} . \label{eq:3}
\end{gather}
Consider a cut $\partial_H A$ of size $\leq \al M$ that we want to
preserve, and let $C \subseteq E$ be the edges in this
cut. 
The probability that an edge in $C$ is contracted on this iteration is
\[ \f{w(C)}{w(E')} ~~\stackrel{(\ref{eq:3})}{\le}~~ \f{w(C)\cd 2h}{OPT_H
    \cd |V'|} ~~\le~~ \f{\al M \cd 2h}{h M\cd |V'|} ~~\le~~ \f{2\al}{|V'|}.\]

Thus, the probability that $C$ survives for all iterations is at least
\[ \prod_{r=h+1}^n \lp1-\f{2\al}r\rp \ge \f{1}{n^{2\al}} .\]
Repeating the algorithm $O(n^{2\al}\logn)$ times produces all $\al
M$-cuts w.h.p., as desired. 
(The bound also holds for non-integer values of $\alpha$, 
using generalized binomial coefficients~\cite{knuth2011art,KS96}.)
Observe that the algorithm does not need to know $OPT_H$ or $M$.
\EP

\subsection{Proofs from \S\ref{sec:extremal-bounds}}
\label{app:small-cuts}

We give the proof of the statements~(2)-(4) of
\Cref{thm:ExtremalCuts}. We first recall the statement of the theorem.

\FewCuts*

\BP[Proof of Theorem~\ref{thm:ExtremalCuts}]
Recall that the proof of statement~(1) is given in
\S\ref{sec:extremal-bounds}.

For statement~(2), we proceed similarly to statement~(1), aiming at a
$k$-cut $\m S^\dag$ with $w_G(\m S^\dag)<2(k-1)$. This time, we greedily
construct an $r_0$-cut $\m S_0$ for some $r_0\in[k-3,k]$ such that
$w(\m S_0)\le r_0\cd(2-(3/5)\g)$.

Let $\m A^2$ be the set of subsets $A\s V$ with $w_G(A)<10/3-\g$; we
initialize $\m S \gets \{V\}$ and $r \gets 1$. In each iteration, our
goal is to increase $r$ by some $r'\in\{2,5\}$ and increase $w_G(\m S)$
by $\le r'\cd (2-(3/5) \g)$.  While $r<k-3$, if there exists a subset
$A\in\m A^2$ that cuts two or more components in $\m S$, then greedily
cut the edges $\pt_G A$ inside $\m S$; $r$ increases by $2$ and
$w_G(\m S)$ increases by at most $10/3-\g\le
2\cd(2-(3/5)\g)$. Otherwise, similarly to case (1), we bucket every
subset $A\in\m A^2$ cutting one component in $\m S$. As long as
$|\m A^2|>2^r+2^{r-1}\cd c_2n^2$, where $c_2$ is the constant in
\Cref{lem:worse}, there exists one component $S\in\m S$ such that there
are $>c_2n^2$ many nonempty buckets $(S,X)$. By \Cref{lem:worse}, there
exist nonempty buckets $(S,X_1),(S,X_2),(S,X_3)$ such that
$(X_1,X_2,X_3)$ $\nicefrac68$-witnesses \Three; take a subset in each
bucket (call them $A_1,A_2,A_3$) and cut the edges
$\pt_G A_1\cup \pt_G A_2\cup \pt_G A_3$ inside $\m S$; $r$ increases by
$5$ and $w_G(\m S)$ increases by at most
$3\cd(10/3-\g)=5\cd(2-(3/5)\g)$.

At the end, we obtain our desired $r_0$-cut $\m S_0$. Finally, to
augment it to a $k$-cut, we proceed identically to case (1), obtaining
cut $\m S^\dag$. Thus,
\begin{align*}
w(\m S^\dag) &\le \tsty (r_0-1)\cd(2-\f35\g) + (k-r_0)\cd (\f{10}3-\g)
 \le \tsty (2-\f35\g)\cdot (k-1) + {O(1)} < 2(k-1) < 2k,
\end{align*}
again using that $k\ge\Om(1/\g)$.

\bigskip
For statement~(3), we proceed similarly, aiming at a $k$-cut $\m S^\dag$
with $w_G(\m S^\dag)<2(k-1)$. This time, we greedily construct an $r_0$-cut
$\m S_0$ for some $r_0\in[k-3,k]$ such that
$w(\m S_0)\le r_0\cd(2-(1/2)\g)$.

Let $\m A^2$ be the set of subsets $A\s V$ with $w_G(A)<4-\g$; we
initialize $\m S$ and $r$ identically to case (1). In each iteration,
our goal is to increase $r$ by some $r'\in\{2,5\}$ and increase
$w_G(\m S)$ by $\le r'\cd (2-(1/2) \g)$.  While $r<k-3$, if there exists
a subset $A\in\m A^2$ that cuts two or more components in $\m S$, then
greedily cut the edges $\pt_G A$ inside $\m S$; $r$ increases by $2$ and
$w_G(\m S)$ increases by at most $4-\g\le 2\cd(2-(1/2)\g)$. Otherwise,
similarly to case (1), we bucket every subset $A\in\m A^2$ cutting one
component in $\m S$. As long as $|\m A^2|>2^r+2^{r-1}\cd cn^{3-1/4}$,
where $c$ is the constant in \Cref{thm:extremal7}, there exists one
component $S\in\m S$ such that there are $>cn^{3-1/4}$ many nonempty
buckets $(S,X)$. By \Cref{thm:extremal7}, there exist nonempty buckets
$(S,X_1),(S,X_2),(S,X_3)$ such that $(X_1,X_2,X_3)$
$\nicefrac78$-witnesses \Three; take a subset in each bucket (call them
$A_1,A_2,A_3$) and cut the edges $\pt_G A_1\cup \pt_G A_2\cup \pt_G A_3$
inside $\m S$; $r$ increases by $6$ and $w_G(\m S)$ increases by at most
$3\cd(4-\g)=6\cd(2-(1/2)\g)$.

At the end, we obtain our desired $r_0$-cut $\m S_0$. Finally, to
augment it to a $k$-cut, we proceed identically to case (1), obtaining
cut $\m S^\dag$. Thus,
\begin{align*}
w(\m S^\dag) &\le \tsty (r_0-1)\cd(2-\f12\g) + (k-r_0)\cd (4-\g)
     \le (2-\f12\g)(k-1) + {O(1)} < 2(k-1) < 2k.
\end{align*}
using that $k\ge\Om(1/\g)$.

\bigskip
For statement~(4), we again proceed similarly, aiming at a $k$-cut
$\m S^\dag$ with $w_G(\m S^\dag)<2(k-1)$. This time, we greedily construct
an $r_0$-cut $\m S_0$ for some $r_0\in[k-3,k]$ such that
$w(\m S_0)\le r_0\cd(2-(3/7)\g)$.

Let $\m A^3$ be the set of subsets $A\s V$ with $w_G(A)<14/3-\g$; we
initialize $\m S$ and $r$ identically to case (1). In each iteration,
our goal is to increase $r$ by some $r'\in\{3,7\}$ and increase
$w_G(\m S)$ by $\le r'\cd (2-(3/7) \g)$.  While $r<k-3$, if there exists
a subset $A\in\m A^3$ that cuts \emph{three} or more components in
$\m S$, then greedily cut the edges $\pt_G A$ inside $\m S$; $r$
increases by $3$ and $w_G(\m S)$ increases by at most
$14/3-\g\le 3\cd(2-(3/7)\g)$. Unlike cases (1) and (2), we have to
separately handle the case when a subset cuts exactly two components in
$\m S$; we will do this next.

For arbitrary subsets $A,S\s V$, let us define
$w_S(A):=(2/M)\cd w(\pt_GA\cap E[S])$, i.e., $2/M$ times the weight of
the cut $\{A\cap S,S\bs A\}$ inside $G[S]$; note that if $S_1,S_2\s V$
are disjoint, then $w_{S_1}(A)+w_{S_2}(A) \le w_G(A)$ for all $A\s
V$. First, suppose that $A$ cuts exactly two components
$S_1,S_2\in\m S$, and that $w_{S_1}(A)\ge 8/3-(4/7)\g$. Then,
\[\tsty w_{S_1}(A)+w_{S_2}(A) \le w_G(A) \implies w_{S_2}(A) \le
  w_G(A)-w_{S_1}(A)\le (\f{14}3-\g)-(\f83-\f47\g)=2-\f37\g.\] In this
case, we greedily cut the edges in $\pt_GA\cap E[S]$; $r$ increases by
$1$ and $w_G(\m S)$ increases by at most $1\cd(2-(3/7)\g)$. Therefore,
we can assume that for each $A$ cutting exactly two components
$S_1,S_2\in\m S$, we have $w_{S_1}(A),w_{S_2}(A)\le 8/3-(4/7)\g$.

For each subset $A\in\m A^3$ that cuts exactly two components
$S_1,S_2\in\m S$ with intersections $A\cap S_1=X_1,A\cap S_2=X_2$, add
$A$ to a bucket labeled with the pair of pairs
$((S_1,X_1),(S_2,X_2))$. By the same observations before, each bucket
has size $\le2^{r-2}$. Now, suppose there are $>2^{r-2}\cd4n^2$ many
subsets $A\in\m A^3$ that cut exactly two components in $\m S$. Then,
there are $>4n^2$ many nonempty buckets, which means that
$>\sr{4n^2}=2n$ many distinct tuples $(S,X)$ are present as a pair in a
nonempty bucket. In other words, there are $>2n$ pairs $(S,X)$ such that
there exists subset $A\in\m A^3$ with $A\cap S=X$. Following case (1),
we conclude that there exist $A_1,A_2\in\m A^3$ such that $A_1\cap S$
and $A_2\cap S$ cross. Cut the edges
$(\pt_G(A_1)\cup\pt_G(A_2))\cap E[S]$; $r$ increases by $3$ and
$w_G(\m S)$ increases by at most
$2\cd(8/3-(4/7)\g) \le 3\cd(2-(3/7)\g)$.

Therefore, we can assume that there are $\le 2^{r-2}4n^2$ many subsets
$A\in\m A^3$ that cut exactly two components in $\m S$. For the subsets
$A\in\m A^3$ cutting exactly one component in $\m S$, we bucket
identically to cases (1) and (2). As long as
$|\m A^3|>2^r + 2^{r-2}4n^2 + 2^{r-1}\cd cn^{4-1/4}$, where $c$ is the
constant in \Cref{thm:extremal}, there exists one component such that
there are $>cn^{4-1/4}$ many nonempty buckets $(S,X)$. By
\Cref{thm:extremal}, there exist nonempty buckets
$(S,X_1),(S,X_2),(S,X_3)$ such that $(X_1,X_2,X_3)$ $1$-witnesses
\Three; take a subset in each bucket (call them $A_1,A_2,A_3$) and cut
the edges $\pt_G A_1\cup \pt_G A_2\cup \pt_G A_3$ inside $\m S$; $r$
increases by $7$ and $w_G(\m S)$ increases by at most
$3\cd(14/3-\g)=7\cd(2-(3/7)\g)$.

At the end, we obtain our desired $r_0$-cut $\m S_0$. Finally, to
augment it to a $k$-cut, we proceed identically to cases (1) and (2),
obtaining cut $\m S^\dag$. Thus,
\begin{align*}
w(\m S^\dag) &\le \tsty (r_0-1)\cd(2-\f37\g) + (k-r_0)\cd (\f{14}3-\g)
 \le (2-\f37\g)\cd (k-1) + {O(1)} < 2(k-1) < 2k,
\end{align*}
using $k\ge\Om(1/\g)$. This completes the proof.
\EP

\subsection{Proofs from \S\ref{sec:recursive-algorithm}}\label{appendix:4}
We prove the following lemmas in \S\ref{sec:recursive-algorithm} that are restated below. 
\gain*

\BP We analyze the value of the
expression of the left hand side
$D := \f{\ell-d(w)}{\ell - (1.75+\Th(\g))}$ for different values of $w$.
\begin{itemize}
\item $w \leq 3 - \gamma$: Since $d(w) = 1$ and $\ell \geq 2$, so $D \geq 1 - O(\gamma)$. 
\item $3 - \gamma \leq w \leq 4 - \gamma$: Since $d(w) = 3 - \nicefrac14$ and $\ell \geq 3$, $D \geq \nicefrac{0.25}{1.25} - O(\gamma) = \nicefrac{1}{5} - O(\gamma)$.
\item $4 - \gamma \leq w \leq 14/3 - \gamma$: Since $d(w) = 4 - \nicefrac14$ and $\ell \ge 4$, $D \geq \nicefrac{0.25}{2.25} - O(\gamma) = \nicefrac{1}{9} - O(\gamma)$.
\item $w > 14/3 - \gamma$: Since $\ell$ is the above the line defined
  in~\eqref{eq:line}, we have
  $\ell \geq (\nicefrac{2s}{k} - 2 - O(\gamma)) w + (8 -
  \nicefrac{6s}{k})$. Now we have
  $s \geq (1.75 + \Theta(\gamma))k$, and $w > \nicefrac{14}{3} - \gamma$, we
  have $\ell \geq 5$.

  Hence, let us consider the case where the integer $\ell = 5 + j$ for
  $j \in \Z_{\geq 0}$. Moreover, the value of $D$ is the smallest when
  $w$ is as large as possible, so we can imagine that~(\ref{eq:line}) is
  tight. Hence,
  \[ \textstyle (5+j) = (2s/k - 2 - O(\g))w + (8-6s/k) \implies w =
    \frac{\nicefrac{6s}{k} + (j-3)}{2(\nicefrac sk-1) - O(\g)}. \]
  Moreover since $d(w) = w$ for these settings of $w$,
  \begin{align*}
    D &= \f{\ell - w}{\ell - 1.75 + \Theta(\g)} = \frac{(5+j) - \frac{\nicefrac{6s}{k}
        + (j-3)}{2(\nicefrac sk-1) - O(\g)}}{(5+j) - 1.75 + \Theta(\g)} \\
      &= \frac{4z + j(2s-k)}{6.5z + 4.875k + j(2s-2k)}(1 - O(\g)).
  \end{align*}

  Since $s \geq 1.75k$, 
  the final expression is smallest when
  $j = 0$, which gives us the first inequality in the lemma. The second
  inequality is immediate since $k \leq k_0$. \qedhere
\end{itemize}
\EP

\pot*

\BP Recall the definitions of $z(k,s)$ from (\ref{eq:z}), $\Phi(k,s)$ from (\ref{eq:pot}), and $g_{k,s}(w)$ from (\ref{eq:g}).
Define $f(t) := \min \lp \f19, \f{4t}{6.5t+4.875k_0}(1 - \Th(\gamma)) \rp$, the term to be integrated in $\Phi(k,s)$. Since $f(t)\le1$ for all $t\ge0$,
\[ \Phi(k,s) \le \int_{t=0}^{z(k,s)}1 = z(k,s) = s-(1.75+\Th(\g))k\le s ,\]
proving the first item. Moreover,
\[ \Phi(k,s)-\Phi(k,s-1) = \int_{t=z(k,s-1)}^{z(k,s)}f(t) \le \int_{t=z(k,s-1)}^{z(k,s)}1 = z(k,s)-z(k,s-1)=1 ,\]
proving the second.
Since
\[ z(k, s) \leq z(k-1, s-1) \leq z(k - 1, s), \] and the potential
$\Phi(k,s)$ integrates a nonnegative function $f(t)$ from $0$ to
$z(k, s) \geq 0$, so it is monotone in $z(k, s)$. This proves the third item.

For the fourth item, note that $\ell \geq 2$ and $\Phi(k - 1, s - \fval) \leq \Phi(k, s)$. 
If $z(k - 1, s - \fval) \geq 0$, 
\begin{align*}
\Phi(k,s) - \Phi(k-1,s-\fval) &= \int_{t=z(k-1, s-\ell)}^{z(k,s)} f(t) dt \\
&\leq (z(k, s) - z(k-1, s-\ell)) \cdot f(z(k, s)) \\
&= (\ell - (1.75 +\Th(\g))) \cdot f(z(k, s)),
\end{align*}
where the inequality uses the fact that $f$ is monotone. 
By \Cref{lem:gain}, the last term is bounded by $\ell - d(w)$ as desired. 

In the case $z(k - 1, s - \fval) < 0 \iff z(k, s) < \ell - (1.75 + \Theta(\gamma))$, 
\begin{itemize}
\item If $\Phi(k, s) \leq 1$, then $\Phi(k, s) \leq 1 + \ell - d(w)$ since $\ell \geq d(w)$. 
\item If $\Phi(k, s) > 1$, it implies $z(k, s) \geq 9 \Phi(k, s) > 9$, and $z(k - 1, s - \fval) < 0 \Leftrightarrow z(k, s) \leq \ell - (1.75 + \Theta(\gamma))$ implies $\ell \geq 10$ and $d(w) \leq \nicefrac89 \cdot \ell$, so
\[
\Phi(k, s) \leq \frac{1}{9} \cdot z(k, s) \leq \frac{1}{9} \cdot (\ell - 1.75) \leq \ell - d(w).
\]
\end{itemize}

Lastly, for the fifth item, we compute
\[ g_{k,s}\inv (\el) = \f{k}{2s-2k}\el + \f{6s-8k}{2s-2k} \le \f{k}{2(1.75+\Th(\g))k-2k}\el+3 \le \f{1}{1.5}\el+3 .\]
On the other hand,
\[ s-\Phi(k,s) \ge s-\int_{t=0}^{z(k,s)}\f19 = s-\f19(s-1.75k-\Th(\g))\ge \f89s.\]
Therefore, $g_{k,s}\inv(s)\le s-\Phi(k,s)+3$, as desired.
\EP


\end{document}